# In Medium Nucleon Structure Functions, SRC, and the EMC effect

Study the role played by high-momentum nucleons in nuclei

**A proposal to Jefferson Lab PAC 38, Aug. 2011**

O. Hen (contact person), E. Piasetzky, I. Korover, J. Lichtenstadt, I. Pomerantz, I. Yaron, and R. Shneor
Tel Aviv University, Tel Aviv, Israel

M. Amarian, R. Bennett, G. Dodge, L.B. Weinstein (spokesperson)
Old Dominion University, Norfolk, VA, USA

G. Ron
Hebrew University, Jerusalem, Israel

W. Bertozzi, S. Gilad (spokesperson), A. Kelleher and V. Sulkosky
Massachusetts Institute of Technology, Cambridge, MA, USA

B.D. Anderson, A.R. Baldwin, and J.W. Watson
Kent State University, Kent, OH, USA

J.-O. Hansen, D.W. Higinbotham, M. Jones, S. Stepanyan, B. Sawatzky, and S. A. Wood (spokesperson), J. Zhang
Thomas Jefferson National Accelerator Facility, Newport News, VA, USA

G. Gilfoyle
University of Richmond, Richmond, VA, USA

P. Markowitz
Florida International University, Miami, FL, USA

A. Beck and S. Maytal-Beck
Nuclear Research Center Negev, Beer-Sheva, Israel

D. Day, C. Hanretty, R. Lindgren, and B. Norum
University of Virginia, Charlottesville, VA, USA

J. Annand, D. Hamilton, D. Ireland, K. Livingston, D. MacGregor, D. Protopopescu, and B. Seitz
University of Glasgow, Glasgow, Scotland, UK

F. Benmokhtar
Christopher Newport University, Newport News, VA, USA

M. Mihovilovic and S. Sirca
University of Ljubljana, Slovenia

J. Arrington, K. Hafidi and X. Zhan
Argonne National Lab, Argonne, IL, USA

W. Brooks, and Hayk Hakobyan
Universidad Técnica Federico Santa María

Theoretical support:

M. Sargsian
Florida International University, Miami, FL, USA

W. Cosyn, and J. Ryckebusch
Ghent University, Ghent, Belgium

G. A. Miller
University of Washington, Seattle, WA, USA

C. Ciofi degli Atti, L.P. Kaptari, and C.B. Mezzetti
INFN and Department of Physics, University of Perugia, Perugia, Italy

M. Strikman
Pennsylvania State University

L. Frankfurt
Tel Aviv University, Tel Aviv, Israel

A. Accardi and W. Melnitchouk
Thomas Jefferson National Accelerator Facility, Newport News, VA, USA

Contents

**Abstract**

# Abstract

We propose to measure semi-inclusive deep inelastic scattering (DIS) off the deuteron by "tagging" the DIS scattering with high momentum recoiling protons or neutrons emitted at large angle relative to the momentum transfer.

While the EMC effect has been observed many times, there is no generally accepted explanation of its origin. Many theoretical models predict that the EMC effect is due to the modification of the nucleon structure functions in the nuclear medium and that this modification increases with nucleon virtuality ($v = (p^\mu)^2 - m^2$). In addition, recent experimental results indicate that most of the EMC effect stems from DIS scattering on high momentum (*i.e.*, high virtuality) nucleons in the nucleus. This proposed measurement on the deuteron will clarify the relationship between modification of the structure function and nucleon virtuality, and might have important implications for understanding the EMC effect and its relation to nucleons in Short-Range Correlations (SRC).

We can study DIS events from high-momentum (high virtuality) nucleons by tagging these events with a backward-recoiling partner nucleon using the reaction $d(e,e'N_s)X$. This can be done by properly choosing kinematics that will be allowed by the 12 GeV upgrade of JLab.

The observable we propose to measure is the ratio of high *x*' ($x' = Q^2/2p^\mu q_\mu$ is the equivalent value of Bjorken *x* for scattering from a moving nucleon) to low *x*' DIS scattering from a tagged partner nucleon in deuterium divided by the same ratio for the untagged scattering. This ratio should be sensitive to the modification of the nucleon structure functions in the medium. At low *x*' (0.25 < *x*' < 0.35), any modification of in-medium nucleon structure functions should be small. At high *x*' (0.45 < *x*' < 0.6) much larger effects are expected. We suggest that if this is true, similar enhancement will be visible by performing the traditional EMC-effect measurement of the ratio of heavy to light nuclei (a LOI to measure this is submitted to this PAC by the same group).

The electrons will be detected in Hall C by the SHMS and HMS, simultaneously covering the high and low ranges of *x*'. The recoiling protons and neutrons will be detected by a Large-Acceptance Detector (LAD) covering the backward angles from about 85º to 174º. To achieve the experimental goals at low cost, we propose to use the time-of-flight (TOF) counters of the current CLAS that will not be used in CLAS12. A dedicated scattering chamber with a large and thin backward window will be constructed to allow the protons to reach the LAD.

We request 6 days for setup and calibrations and 34 days for data taking on deuterium for a total of 40 days. Note that a PAC decision is needed now to make sure that the proposed LAD scintillator counters are removed intact when CLAS6 is decommissioned.

# I. Introduction

## I.1 The EMC effect

One of the outstanding questions in nuclear physics is whether the quark structure of nucleons is modified in the nuclear medium. Evidence for nucleon modification can only come from the failure of hadronic models, models incorporating unmodified nucleons and mesons as their fundamental degrees of freedom. While the vast majority of nuclear physics experiments can be explained using hadronic models, deep inelastic scattering (DIS) measurements (see Fig. 1) of the ratio of per-nucleon cross sections of nucleus $A$ to deuterium cannot. These experiments typically measure a ratio of about 1 at $x = 0.3$, decreasing linearly to a minimum at around $x = 0.7$ [1–7]. This minimum depends on $A$ and varies from about 0.94 for $^4$He to about 0.83 for $^{197}$Au (see Fig. 2). This observation is known as the EMC effect. A comprehensive review of the EMC effect can be found in [8, 9] and the references therein.

While the EMC and other modifications have been observed, there is no generally accepted explanation of their origin. In general, two classes of explanations have been proposed: 1) the internal structure of the nucleon is modified by the influence of the nuclear medium or 2) there are effects stemming from the many body nuclear medium itself, such as binding, etc. Note that a recent publication indicates that the Coulomb field of the nucleus also plays an important role [10].

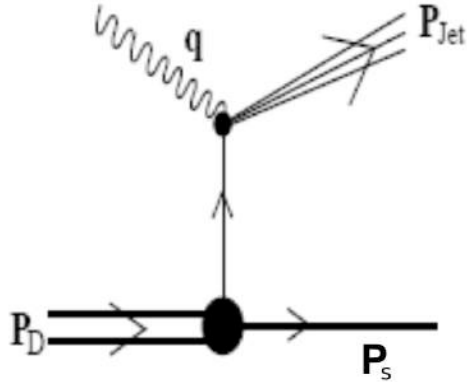

**Figure 1:** Deep inelastic scattering from deuterium showing the momentum transfer and the momentum of the recoil spectator.

There are many models of nucleon modification used to explain the EMC effect [8,9]. We show here as an example a recent work by Kulagin and Petti [11,12] which includes nuclear shadowing, Fermi motion and binding, nuclear pion excess and a phenomenological off-shell correction to bound nucleon structure functions. They assumed that the off-shell correction is proportional to the nucleon virtuality $v = (p^\mu)^2 - m^2$ and parametrized it as a third-order polynomial in $x$. Fig. 2 shows that their calculation describes the data very well over a wide range of $x$ ($x = Q^2/2m\omega$ where $Q^2$ is the four-momentum transfer squared, $\omega$ is the energy transfer and $m$ is the nucleon mass). The offshell correction is required to describe the data.

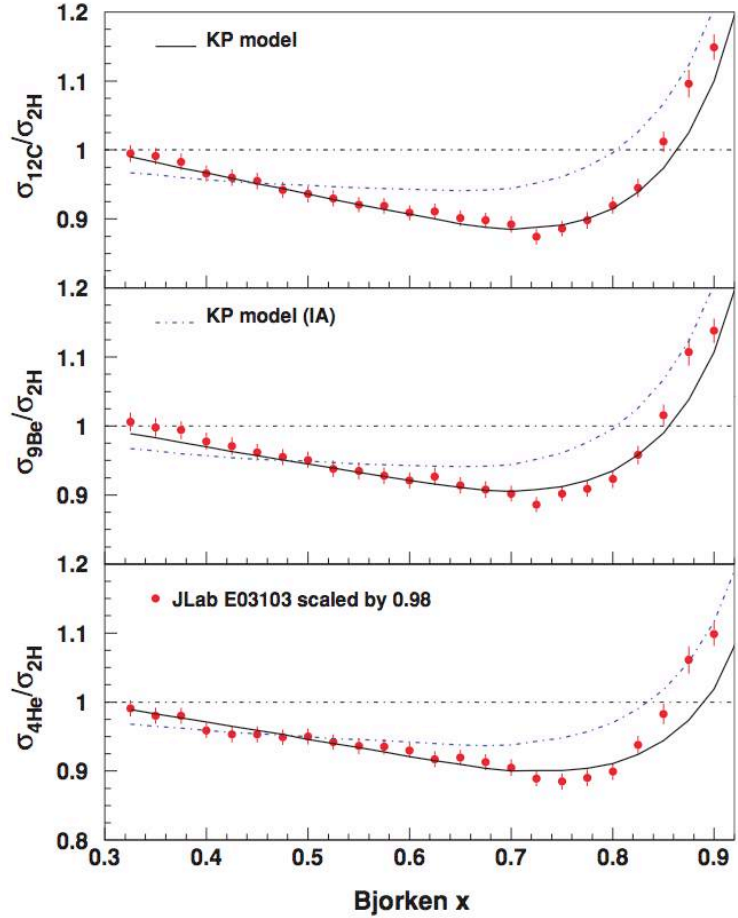

**Figure 2:** The per-nucleon DIS cross section ratio of $^{12}$C, $^{9}$Be, and $^{4}$He to deuterium as a function of $x$. The points are from Seely [7] and the curves are from Kulagin and Petti [11,12]. The solid curve shows the full model and the dot-dashed curve shows the result with no off shell correction.

Recent published measurements of the EMC effect on light nuclei at JLab [7] show that the EMC effect does not depend directly on the atomic mass $A$ or the average nuclear density, as was previously assumed by some models. Beryllium is the most significant outlier (see Fig. 3). The authors claim, "The data … suggest that the nuclear dependence of the quark distributions may depend on the local nuclear environment". These data suggest that nucleon modification increases with local nuclear density. This implies that we should compare the EMC effect to other density-related nuclear phenomena such as short range correlations.

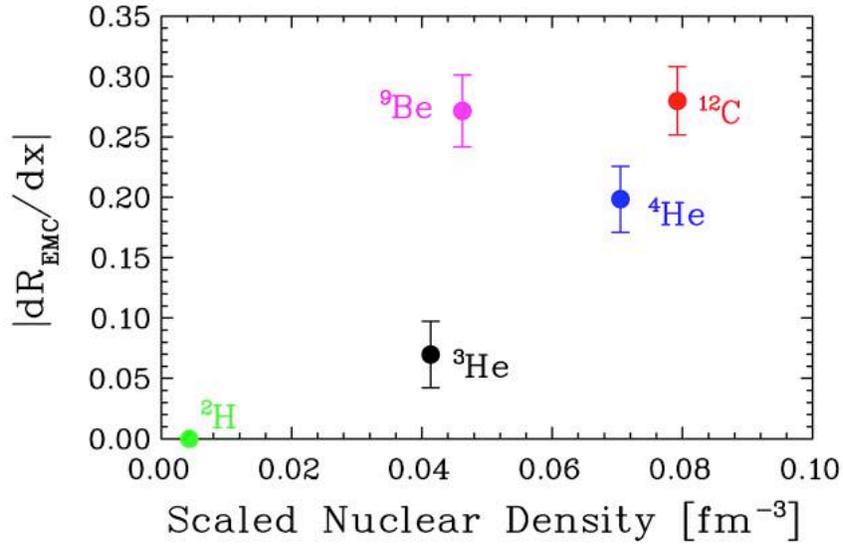

**Figure 3:** The slope of the EMC ratios for light nuclei from Seely [7] plotted vs the scaled average nuclear density.

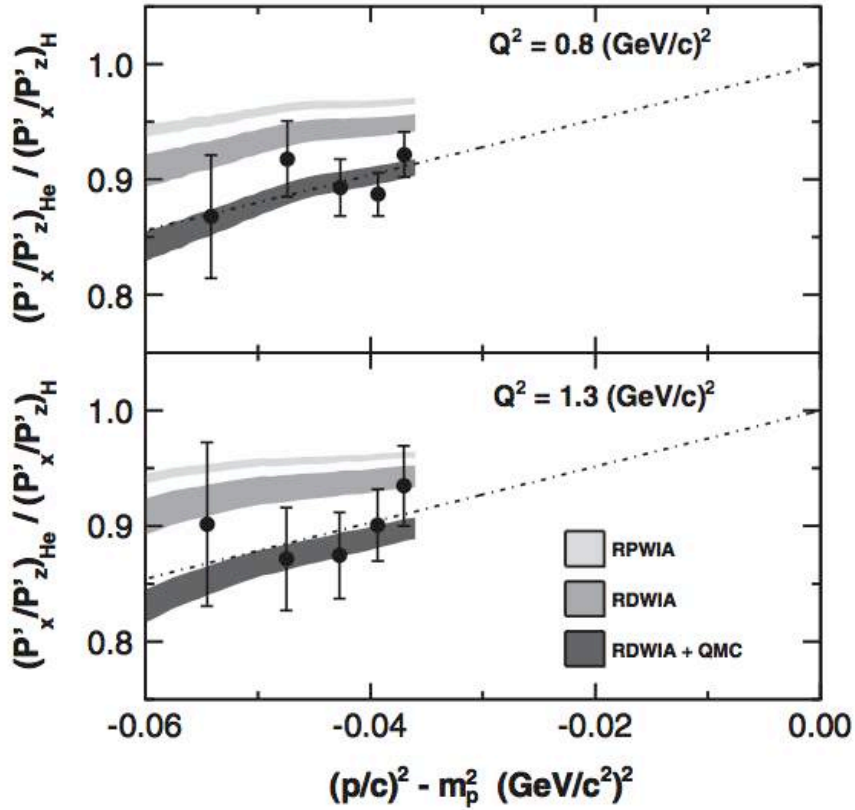

**Figure 4:** The double ratio of proton polarization in the *x'* and *z'* directions for $^4\text{He}(\vec{e},e'\vec{p})^3\text{H}$ relative to $\text{H}(\vec{e},e'\vec{p})$ plotted versus nucleon virtuality showing deviation from the free nucleon for $Q^2 = 0.8$ and $1.3$ GeV$^2$ [13].

Another recent JLab measurement also seems to indicate that the medium modifications of the proton electromagnetic form factors increase with nucleon virtuality [13]. Proton recoil polarization was measured in the quasielastic $^4\text{He}(\vec{e},e'\vec{p})^3\text{H}$ reaction at $Q^2 = 0.8$ GeV$^2$ and 1.3 GeV$^2$. The polarization-transfer coefficients were found to differ from those of the $\text{H}(\vec{e},e'\vec{p})$ reaction as the virtuality of the proton increases (see Fig. 4). Note that this experiment only explored relatively small nucleon virtualities, since the proton momenta were significantly below the Fermi momentum.

**I.2 Short Range Correlation (SRC) in nuclei**

Only about 60-70% of nucleons in nuclei are in single-particle mean-field orbitals. Some nucleons are in long-range correlated pairs and the rest of the nucleons are in short-range correlated (SRC) *NN* pairs. These SRC pairs are characterized by a large relative momentum and small center-of-mass momentum, where large and small are relative to $k_F$, the Fermi momentum of heavy nuclei [14–16]. In other words, when a nucleon belongs to an SRC pair, its momentum is balanced by one other nucleon, not by the $A-1$ other nucleons.

We have learned a large amount about these correlated pairs in the last decade from experiments at Jefferson Lab [17-24] and BNL [25-28]:

● The probability for a nucleon to belong to an SRC pair ranges from 5% in deuterium to about 25% in nuclei such as carbon and iron;

● The threshold momentum of nucleons in SRC pairs is $p_{thresh} = 275 \pm 25$ MeV/c;

● The momentum distribution for $p > p_{thresh}$ is the same for all nuclei, only the magnitude varies. This magnitude is expressed as a scale factor, $a_{2N}$;

● Almost all nucleons with momenta greater than $p_{thresh}$ are part of *NN*-SRC ($92 \pm 18\%$);

● These SRC pairs move inside the nucleus with c.m. motion of σ~0.14 GeV/c;

● The *NN*-SRC consists of about 90% *np* pairs, and 5% each *pp* and *nn* pairs;

● The tensor force dominates *NN*-SRC for pair relative momenta $0.3 < p_{rel} < 0.5$ GeV/c

● 80% of the kinetic energy (momentum) of all the nucleons in the nucleus is carried by members of the *NN*-SRC (which are only 20% of the nucleons).

A pie chart that represents our 'standard' picture of $^{12}$C short range structure is shown in Fig 5.

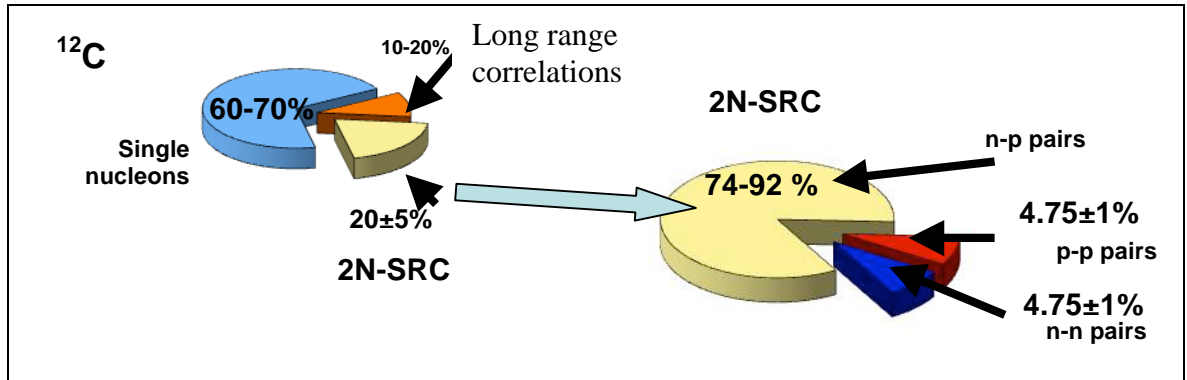

**Figure 5:** The short distance structure of $^{12}$C as deduced from recent measurements.

### I.3 SRC and the EMC effect

A recent paper [29] found that the size of the EMC effect in a given nucleus is closely correlated with the probability for a nucleon in that nucleus to belong to an *NN* SRC pair (see Fig. 6). The dependence of the EMC effect on high momentum nucleons was first proposed in [30]. This strongly suggests that the EMC effect is due to high momentum nucleons in nuclei. Since almost all high-momenta nucleons in nuclei belong to SRC nucleon pairs, we can select the nucleons on which we observe the EMC effect by detecting their SRC partners that recoil backwards in coincidence with the scattered electrons.

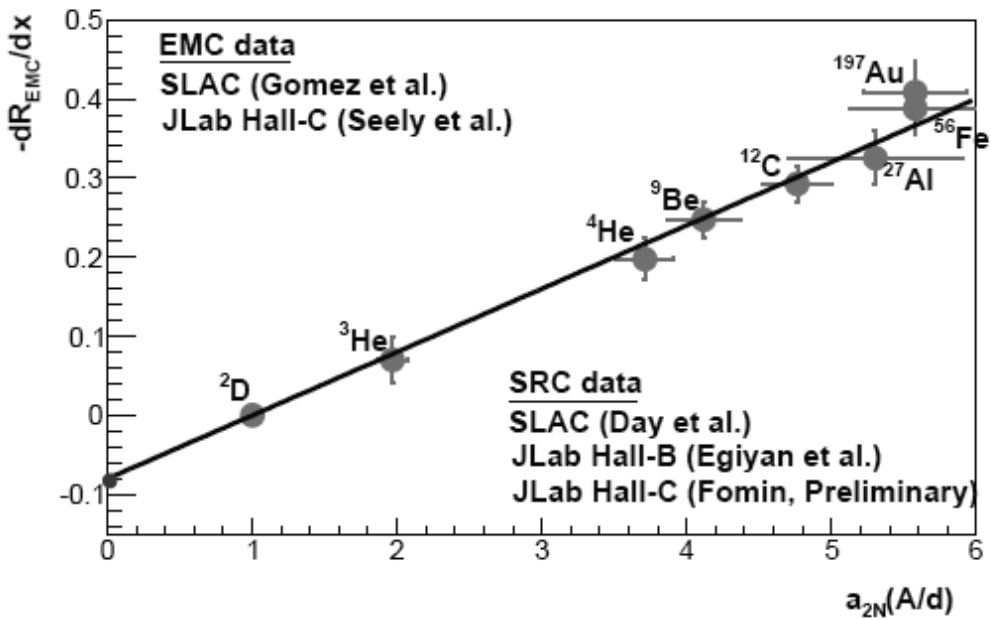

**Figure 6:** The negative of the EMC slope plotted vs. the relative probability that a nucleon belongs to an *NN* SRC pair for a variety of nuclei (see details in [29]).

## I.4 DIS off nuclei in coincidence with high momentum recoil nucleons

If the EMC effect is predominantly associated with 2N-SRC pairs in nuclei, then the per-nucleon ratio of the tagged DIS cross sections for $d$ and nucleus $A$ (the "tagged" EMC ratio) should be almost independent of $x$ and larger than unity. This is because, in both nuclei, the spectator backward nucleon tags the reaction so that the electron is scattering from a high momentum forward-going nucleon. Thus, the electron is scattering from nucleons with the same momenta and virtuality in both nuclei. If the nucleon modification and hence the EMC effect depend on the virtuality of the struck nucleon rather than on the nuclear density, the per-nucleon cross section ratio (the EMC ratio) of the two nuclei should be independent of $x$. The magnitude of the ratio should equal the relative probabilities for a nucleon to belong to a SRC pair in those two nuclei.

Fig. 7 shows a very preliminary analysis of CLAS eg2 5 GeV data. Note that the data are not corrected for final state interactions, radiative effects, acceptance effects, etc. The bottom plot shows the EMC ratio of the per-nucleon DIS $(e,e')$ cross sections for C and deuterium. It shows the typical linear decrease in the ratio from $x = 0.3$ to $x = 0.6$. Note that we allow $Q^2$ values as low as 1 GeV$^2$ in order to increase statistics. However, higher twist effects should cancel in the EMC ratio. The top plot shows the tagged EMC ratio, requiring that a high momentum proton ($p > 0.3$ GeV/c) be detected at an angle greater than 120 degrees from the momentum transfer. Although the statistics are marginal, the results are consistent with being constant with $x$ and the value of the ratio is about 6, slightly larger than the expected ratio of $a_{2N}$(C/d) = 4.8 ± 0.4 [29].

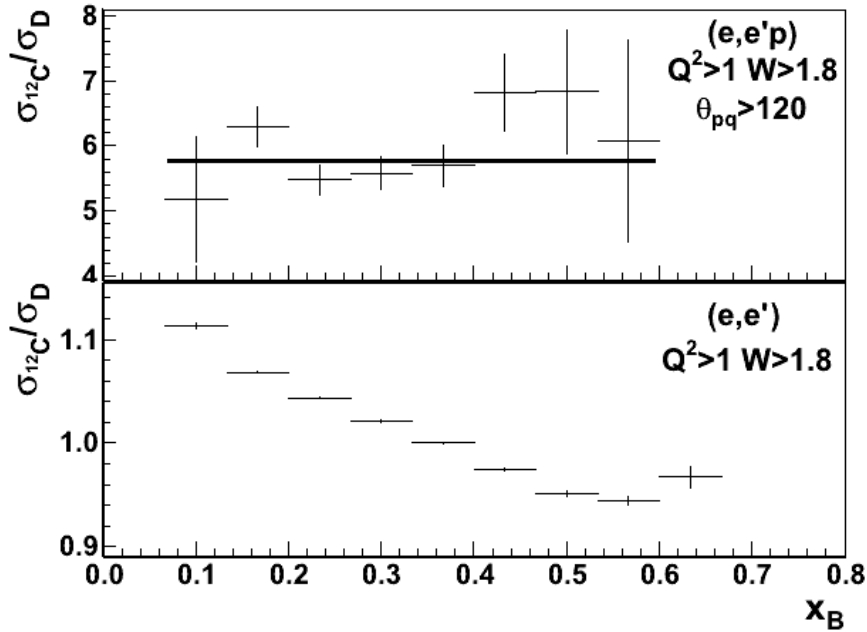

**Figure 7: Preliminary and not for release.** The per-nucleon DIS cross section in $^{12}$C divided by the same quantity for the deuteron. Bottom: untagged inclusive cross section; top: the cross section tagged by a high momentum, $p > 0.3$ GeV/c, backward proton. The data are a preliminary analysis of CLAS data by O. Hen

While it would be premature to draw quantitative conclusions from this preliminary data, the tagged EMC ratio is clearly very different from the untagged data and agrees with our simple SRC-correlated virtuality-dependent EMC effect idea.

**I.5 DIS off the Deuteron in coincidence with high momentum recoil nucleons**

Due to the lack of a free neutron target, the EMC measurements used the deuteron as an approximation to a free proton plus neutron system and measured the ratio of inclusive DIS on nuclei to that of the deuteron. This seems like a reasonable approximation since the deuteron is loosely bound ($\approx$ 2 MeV) and the average distance between the nucleons is large ($\approx$ 2 fm). But the deuteron is not a free system. We define the In-Medium Correction (IMC) effect as the ratio of the DIS cross section per nucleon bound in a nucleus relative to the free *pn* pair cross section (as opposed to the EMC effect which uses the ratio to deuterium).

The deuteron IMC effect can be extracted from the data in Fig. 6. If the EMC/IMC effect and the SRC scaling factor both stem from the same cause, then the IMC effect and the SRC scaling factor, $a_{2N}(A/d)$, will both vanish at the same point. The value $a_{2N}(A/d) = 0$ is the limit of free nucleons with no SRC. Extrapolating the best fit line in Fig. 6 to $a_{2N}(A/d) = 0$ gives a *y*-intercept of $dR_{EMC}/dx = -0.079 \pm 0.006$. The difference between this value and the deuteron EMC slope of 0 is the deuteron IMC slope. Following [29] the expected IMC effect on a deuteron at *x*=0.6 is about:

$$\frac{\sigma_d}{\sigma_p + \sigma_n} = 1 - (0.079 \pm 0.006)(0.6 - 0.31 \pm 0.04) \approx 0.975$$

where 0.079 is the expected EMC slope for the ratio of a free proton plus neutron to deuterium, and (0.6 – 0.31) is the difference in *x* from 0.31 (where the EMC effect is zero) to *x* = 0.6. Thus we expect that the deuteron DIS cross section at *x* = 0.6 is 2.5% smaller than the sum of the proton plus neutron DIS cross sections at *x* = 0.6.

What will happen if we perform the DIS scattering off a nucleon with large momentum at the tail of the deuteron wave function? If the IMC depends on the virtuality of the interacting nucleon, defined via the kinematics of the spectator, we can expect the IMC effect to be substantially enhanced compared to the effect measured in the inclusive measurements.

If the dominant contribution to the IMC effect is due to the high momentum tail nucleons, then this 2.5% effect is due to the 5% of nucleons with $p > p_{thresh} \approx 275$ MeV/c. Therefore we expect that the modification of these high-momentum nucleons to be of the order:

$$\frac{\sigma_p^*}{\sigma_p} \approx \frac{\sigma_n^*}{\sigma_n} \approx \frac{2.5\%}{5\%} \approx 50\%$$

where $\sigma_p^*$ and $\sigma_n^*$ are the medium-modified high-virtuality DIS cross sections and the probability of finding a nucleon with $p >$ 275 MeV/c in the deuteron is about 5%. Note that this is a qualitative argument that neglects effects such as Fermi motion. However, it should indicate the order-of-magnitude of the effect.

Deuterium is the optimal system in which to study the dependence of the nucleon structure on the nucleon virtuality. The probability for a high momentum configuration in the deuterium is rather small relative to heavier nuclei but this configuration can be 'tagged' cleanly by the emission of a fast nucleon to the backward hemisphere. In a simple spectator picture with no FSI the backward moving nucleon is a spectator, does not participate in the DIS process, and allows us to determine the virtuality of the nucleon from which the electron scattered. The effect of FSI will be discussed in section II.3 below.

# II. Theoretical Background

## II.1 The Formalism for inclusive and semi-inclusive DIS

The inclusive cross section for an electron scattering off a free nucleon at rest in the laboratory frame can be expressed in the DIS region in terms of the structure functions $F_1$ and $F_2$:

$$\frac{d^3\sigma}{d\Omega dE'} = \left(\frac{d\sigma}{d\Omega}\right)_{Mott} \cdot \left[\frac{1}{\omega}F_2(x_B,Q^2) + \frac{2}{M}F_1(x_B,Q^2)\cdot \tan^2(\theta_e/2)\right],$$

where the Mott cross section is the cross section to scatter off a point charge, $\theta_e$ is the electron scattering angle in the lab frame, $Q^2$ is the four momentum transfer, and $x$ is the Bjorken scaling variable given by:

$$x = \frac{Q^2}{2M\omega}.$$

The structure functions $F_1$ and $F_2$ are related by $R$, the ratio between the cross sections for longitudinal and transverse scattering. Using the measured values of $R$, $F_2$ can be extracted from the measured cross section:

$$F_2(x,Q^2) = \left[\frac{d^2\sigma/d\Omega dE'}{(d\sigma/d\Omega)_{Mott}}\right] \cdot \left[\frac{\omega\varepsilon(1+R(x,Q^2))}{1+\varepsilon R(X,Q^2)}\right],$$

where the polarization of the virtual photon is: $\varepsilon = [1+2(1+Q^2/(4M^2))\tan^2(\theta_e/2)]^{-1}$.

An important kinematical parameter that defines the DIS region is the total hadronic mass squared which for a free nucleon at rest is:

$$W^2 = M^2 - Q^2 + 2M\omega$$

When scattering off a nucleon in a nucleus, the movement of the nucleon needs to be taken into account. For a free nucleon moving in the lab frame, the Bjorken scaling parameter is

$$x' = \frac{Q^2}{2p_\mu q^\mu},$$

where $p_\mu$ and $q_\mu$ are the four vectors of the struck nucleon and the virtual photon respectively. For a nucleon at rest $p_\mu = (M,\vec{0})$ and $x' = x$.

In a deuteron, $x'$ can be expressed in terms of the measured recoil tagged nucleon:

$$x' = \frac{Q^2}{2[(M_d - E_s)\omega + \vec{p}_s \cdot \vec{q}]},$$

where $E_s$ is the energy of the recoil nucleon and $\vec{p}_s$ and $\vec{q}$ are the three-momenta of the recoil nucleon and virtual photon respectively. Note that $\vec{p}_s \cdot \vec{q} < 0$ since $\theta_{qs} > 90°$ and hence $x' > x$.

The total hadronic mass for a moving nucleon absorbing a virtual photon in deuterium can also be written in terms of the recoil spectator momentum:

$$W'^2 = (q^\mu + p_d^\mu - p_s^\mu)^2$$

where $p_d^\mu = (M_d, \vec{0})$ is the 4-momentum of the deuteron (here we are being a little careless about the difference between covariant and contravariant four-momenta). This gives:

$$W'^2 = m^2 - Q^2 + 2\omega(M_d - E_s) + 2\vec{p}_s \cdot \vec{q} = m^2 - Q^2 + 2\omega(M_d - E_s) + 2p_s\sqrt{Q^2 + \omega^2}\cos(\theta_{sq})$$

where $\theta_{sq}$ is the angle between the virtual photon and the recoil nucleon and $|\vec{q}| = \sqrt{Q^2 + \omega^2}$.

To ensure a DIS process, we require for the moving nucleon:
$Q^2 \geq 2$ GeV$^2$
$W'^2 \geq 4$ GeV$^2$.

Note that, while higher twist effects increase as $Q^2$ decreases, these effects mostly cancel in the ratio of cross sections. That is why EMC cross section ratios are almost independent of $Q^2$ for $2 < Q^2 < 40$ GeV$^2$ [3].

For a deuteron, even in the presence of FSI, the cross section for the semi-inclusive process is given by [34,40]:

$$\frac{d^4\sigma}{dxdQ^2 d\vec{p}_2 d\phi_e} = KS_D(\vec{p}_s, \vec{q})F'(x', \alpha_s, p_T, Q^2),$$

where $F'$ is the in-medium structure function of the nucleon in the deuteron, $K$ is a kinematic factor, and $S_D$ is the distorted deuteron momentum distribution. The essential point is that in tagged DIS, unlike in quasielastic scattering, factorization of the cross section into the nuclear structure function and the distorted momentum distributions is a good approximation.

The deuteron distorted momentum distribution can be expressed as a function of the measured parameters $S_D = S_D(p_s, \theta_{sq}, W', Q^2)$ where the momentum distribution of the deuteron is determined by $p_s$ and the FSI depends primarily on $\theta_{sq}$ and the invariant mass of the outgoing hadrons $W'$.

## II.2 Off shell cross section models

Motivated in part by the EMC effect results, theorists have proposed many different models of offshell nucleons. Melnitchouk, Sargsian and Strikman [34] calculated the change in the nucleon structure function in deuterium for three different models: a Point-Like Configuration (PLC) suppression model, a binding/offshell model, and a rescaling model where the change in quark localization from the deuteron to heavy nuclei is related to a $Q^2$ rescaling. Fig. 8 shows the predictions for the effective proton structure function in nuclei divided by the free $F_{2p}$ for the different models as a function of $\alpha_s = (E_s - p_s^z)/m_s$ and of $x$ where $\alpha_s$ is the light cone variable, and $E_s, p_s^z$ and $m_s$ are the energy, component of the momentum parallel to the virtual photon, and mass of the backward spectator nucleon. At $\theta_{pq} = 180^\circ$, $\alpha_s = 1.5$ corresponds to $p_s^z = -0.4$ GeV/c.

Melnitchouk, Schreiber and Thomas [35,36] calculated the ratio of the bound to free neutron structure functions in a covariant model with relativistic vertex functions which parametrize the nucleon--quark-"diquark" interaction (where "diquark" just refers to a nucleon with one quark knocked out). The parametrization is constrained by fitting to on-shell structure functions. Figure 9a shows that they find much smaller effects in the ratio of $F_{2n}^{eff}/F_{2n}$ than the previous model. The results for a similar model by Gross and Liuti [37,38] are shown in Fig. 9b. Note that they expect a much larger change in the bound nucleon structure function but a much smaller dependence on $x$.

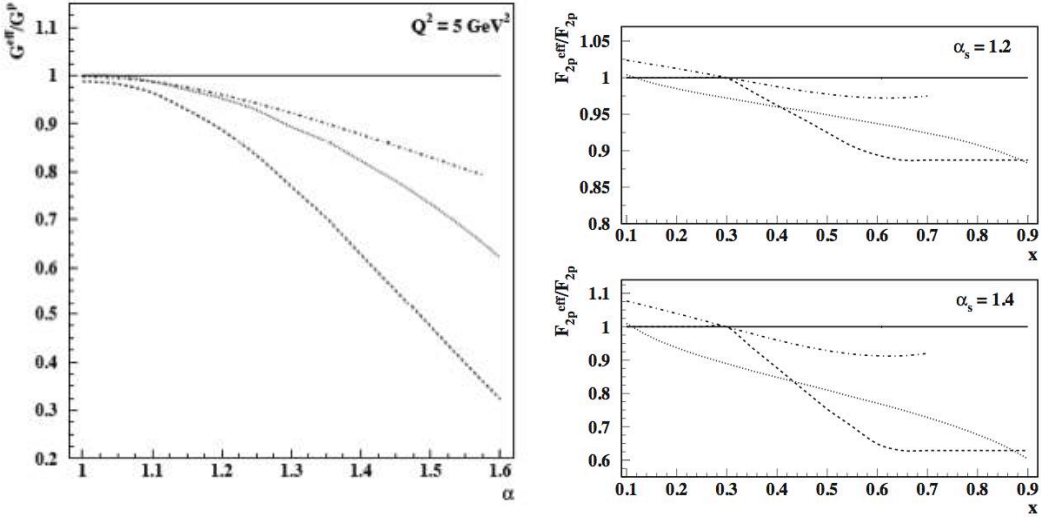

**Figure 8:** (left) The $\alpha_s$ dependence of $F_{2p}^{eff}/F_{2p}$ in deuterium for $x = 0.6$, $Q^2 = 5$ GeV$^2$ and $p_T = 0$ ($\theta_{pq} = 180^\circ$) [34]. (right) The $x$ dependence of $F_{2p}^{eff}/F_{2p}$ for (upper) $\alpha_s = 1.2$ and (lower) $\alpha_s = 1.4$. The dashed line shows the PLC suppression model, the dotted line shows the rescaling model, and the dot-dashed line shows the binding/off-shell.

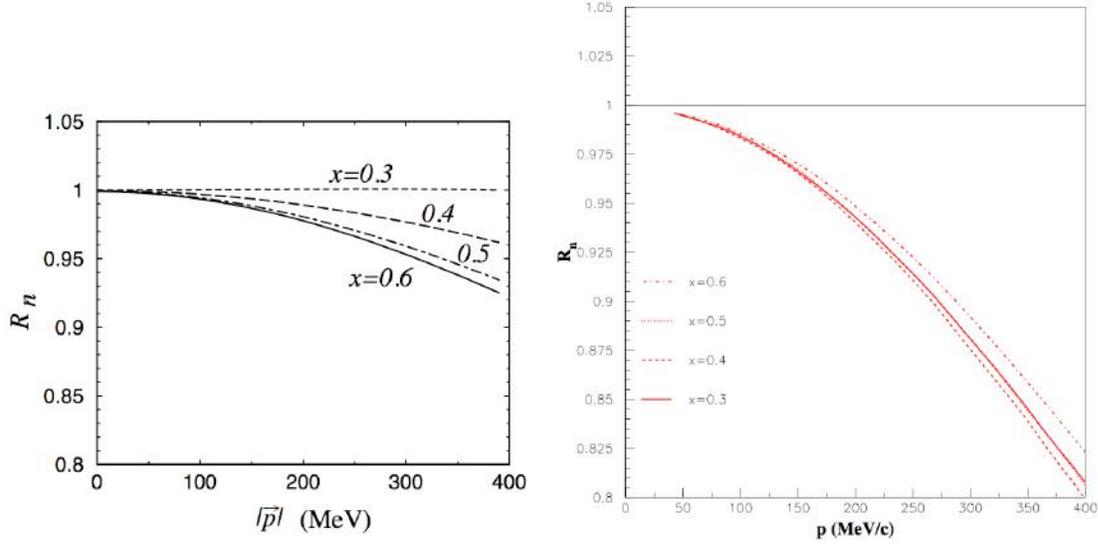

**Figure 9:** The dependence of the ratio $R_n = F_{2n}^{eff}/F_{2n}$ in deuterium for $Q^2 = 5$ GeV$^2$ as a function of spectator proton momentum in (left) the model of Melnitchouk, Screiber and Thomas [36] and (right) the model of Gross and Liuti [37,38] as shown in Ref. [39].

Because the different models predict very different $\alpha_s$, $p_s$ and $x$ dependences, we will measure the cross section as a function of all of those variables. Note also that these model uncertainties, coupled with the absence of free neutrons targets, means that our knowledge of $F_{2n}$ is poor.

## II.3  Final State Interactions (FSI)

FSI are due to the interactions of the recoiling nucleon with the propagating struck nucleon debris formed after the virtual photon absorption by a quark. Note that this is complicated by propagation and hadronization of the struck quark and of the residual system.

While there is no complete theory of FSI in DIS, there are a number of phenomenological models for the deuteron. The magnitude of the FSI in the reaction $d(e,e'p_s)$ has been calculated in several models using PWIA [34], general eikonal approximation as fit to data [40], and with models for the debris-nucleon interaction cross sections [41,42]. There is a general agreement that FSI should be suppressed at backward spectator angles. This agreement is also supported by $d(e,e'p_s)X$ data from CLAS [31]. Figure 10 shows that the PWIA spectator picture describes the data rather well for proton angles larger than 107 degrees relative to the momentum transfer direction (panel a). On the other hand, at angles around $90^0$ (panel b), a large excess of high-momentum protons over the PWIA spectator expectation is observed, which is most likely due to strong final state interactions.

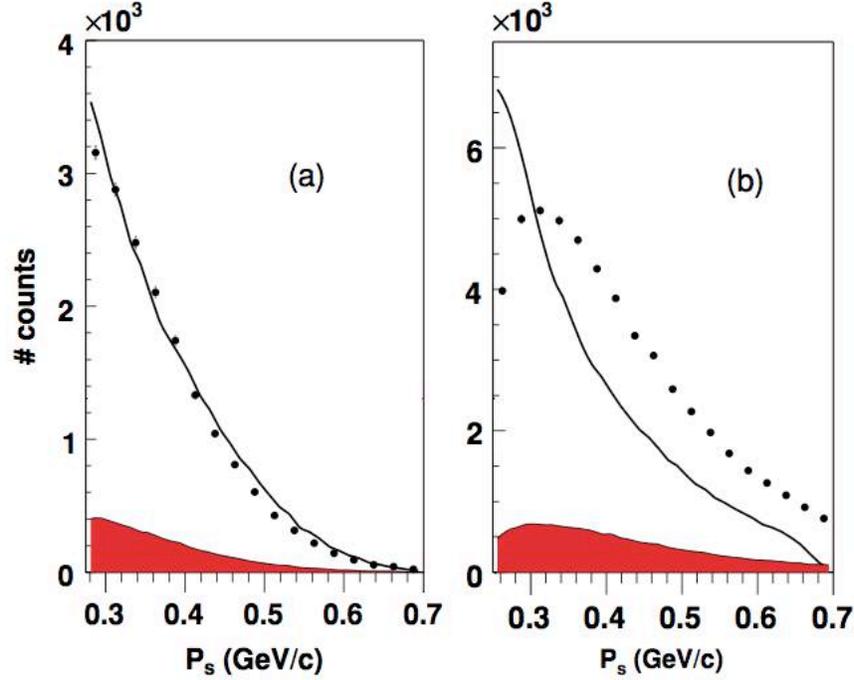

**Figure 10:** Momentum distribution of the recoiling proton in the reaction $d(e,e'p_s)X$ [31]. Data (points) are compared with a PWIA calculation integrated over the experimental acceptance for the range of recoil angle (a) $-1 < \cos\theta_{pq} < -0.3$ and (b) $-0.3 < \cos\theta_{pq} < 0.3$. Events were integrated over all $W$ and $Q^2$.

The different model calculations agree that FSI increase with $W'$ and decrease with $Q^2$. The FSI do not depend strongly on $x'$, thus the ratio of cross sections for two different value of $x'$ is much less sensitive to FSI.

While recent model calculations agree that FSI are smaller for backward recoil nucleons, they strongly disagree about the magnitude of the FSI. Cosyn and Sargsian [40] predict little or no FSI in the backward direction, whereas Ciofi degli Atti and collaborators [41,42] found large FSI even at backward recoil angles (depending on the kinematics).

Therefore it is important to measure ratios of cross sections at different kinematics chosen to maximize or minimize the sensitivity of the ratios to FSI.

To demonstrate the reduction in the sensitivity to FSI we can use the calculation by Ciofi [41,42] which predicts the largest backward angle FSI. According to this calculation, for $p_s = 0.3$ GeV/c, $W' = 2$ GeV$^2$ and $\theta_{pq} = 145°$ the difference between the cross section with and without FSI can be as much as a factor 2 for $x' = 0.27$ and $x' = 0.37$. However, because $p_s$, $W'$ and $\theta_{ps}$ are held constant, the double ratio of the PWIA to the full calculation at these two $x'$ values is unity to better than 1%.

We will measure cross sections over a very wide range of kinematics so that we can study the variation of FSI and other cross section ingredients with $p_s$, $W'$, $x'$, etc.

We will then determine ratios of experimental cross sections measured in kinematics such that FSI, at a given value of the recoil moment of the detected neutron ($p_s$), are minimized ($\theta_{ps} > 110^o$) and relatively constant (fixed $W'$). These ratios at different $x'$ will be used to measure the in medium structure functions of bound nucleons and look for their possible medium modification.

## II.4 Extracting in-medium nucleon structure functions

Ideally one could measure the $d(e,e'N_s)$ cross section and extract from it the corresponding nucleon structure function, but the cross section depends upon: (i) the distorted momentum distribution, which in turn depends on the FSI in a model dependent fashion, and (ii) the in-medium structure function we want to investigate. Fortunately, one can take advantage of the fact that the tagged DIS cross section factorizes into the nucleon structure function times the distorted momentum distribution. Therefore we should choose pairs of kinematic points such that the effects of FSI and the deuteron momentum distribution cancel in the ratio of cross sections. In other words, we should measure the ratio of cross sections at kinematics which have the same $p_s$ and $W'$ but different $x'$. In this case, the cross section ratio should be very sensitive to any distortions of the in-medium nucleon structure functions. Even under these conditions theoretical corrections are needed. However, the corrections and hence their uncertainty will be much smaller than in trying to extract the in-medium structure function directly from the cross section measurements.

The ratio between the $d(e,eN_s)$ cross section at two different $x'$ values, keeping the recoil nucleon kinematics the same, is:

$$\frac{d^4\sigma}{dx_1 dQ^2 d\vec{p}_s} / \frac{d^4\sigma}{dx_2 dQ^2 d\vec{p}_s} = (K_1/K_2)[F_2^*(x_1',\alpha_s,p_T,Q_1^2)/F_2^*(x_2',\alpha_s,p_T,Q_1^2)]$$

Using $x_1' \approx 0.45 - 0.6$ and $x_2' \approx 0.3$ we will measure:

$$[F_2^*(x_1',\alpha_s,p_T,Q_1^2)/F_2^*(x_2',\alpha_s,p_T,Q_1^2)] = \left(\frac{d^4\sigma}{dx_1 dQ^2 d\vec{p}_s}/K_1\right) / \left(\frac{d^4\sigma}{dx_2 dQ^2 d\vec{p}_s}/K_2\right)$$

Integrating over the recoil scattering angle in the range where the FSI is expected to be small, we will compare the measured ratio as a function of $\alpha_s$ to the BONUS results for the free neutron structure function in the $d(e,e'p_s)$ reaction [32,33] and to the measured free proton structure function in the $d(e,e'n_s)$ case.

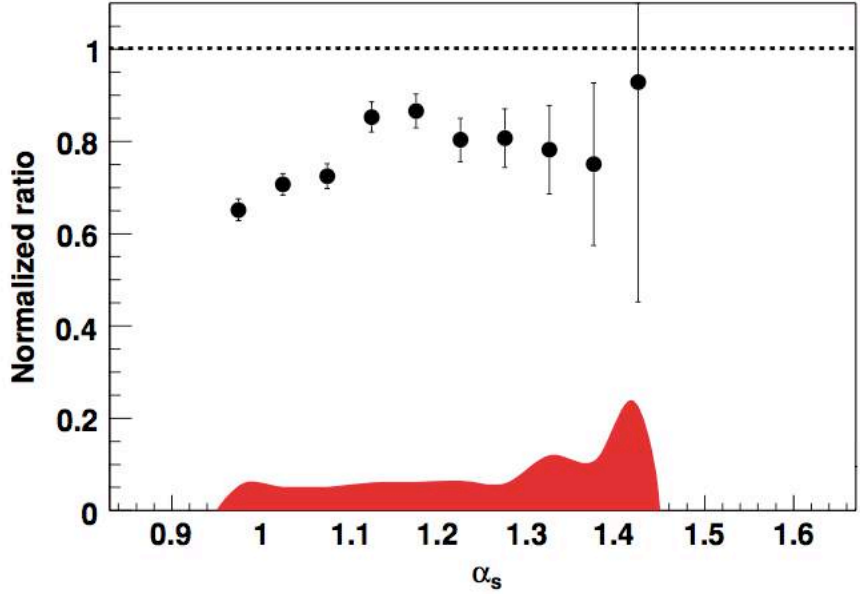

**Figure 11:** Ratio of the extracted off-shell structure function $F_{2n}$ at $x' = 0.55$, $Q^2 = 2.8\,\text{GeV}^2$ to that at $x' = 0.25$, $Q^2 = 1.8\,\text{GeV}^2$, divided by the ratio of free structure functions at those kinematic points. The error bars show the statistical uncertainty; the shaded band indicates the systematic uncertainty. [31]

This ratio was already measured in the CLAS Deeps experiment [31]. Fig. 11 shows the ratio of the extracted in-medium structure function $F_{2n}^{eff}$ at $x' = 0.55$, $Q^2 = 2.8\,\text{GeV}^2$ to that at $x' = 0.25$ $Q^2 = 1.8\,\text{GeV}^2$, divided by the ratio of free structure functions at those kinematic points (substituting $x$ for $x$') for $0.25 \le p_T \le 0.35$ GeV/c plotted versus the spectator nucleon light cone fraction $\alpha_s = (E_s - p_s^z)/m$. In this ratio the effects of the deuteron momentum distribution should cancel. However, FSI do not appear to cancel. This is most apparent at $\alpha_s \approx 1$ ($\theta_{qs} \approx 90^o$) where FSI effects are largest and the ratio is smallest.

For $\alpha_s \ge 1.2$ the measured ratio is consistent with being constant. This disagrees with the more dramatic predictions shown in Figs. 8 and 9. However, due to the limited kinematic flexibility available at 6 GeV, $W$' varied with $\alpha_s$. Therefore the effects of FSI probably also varied with $\alpha_s$.

It is important to remeasure these ratios with enough data over a broad enough kinematic range to understand and control the Final State Interactions correction and other systematic effects.

# III. Details of the proposed measurements

## III.1 The proposed kinematics

The goal of the proposed experiment is to measure the in-medium neutron and proton structure functions and to compare them to the free ones. The obstacles to this are:

1. The deuteron wave function is not well known at the large momentum that we plan to tag the DIS with.

2. The FSI are not well known.

To overcome these obstacles we plan to:

1. Compare measured cross sections to a variety of state of the art calculations over a broad kinematical range in order to check and/or optimize the calculations.
    a. Measure cross sections at $\theta_{ps} \approx 90°$ to study FSI,
    b. Measure cross sections at $\theta_{ps} > 110°$ to minimize FSI,
2. Construct ratios of cross sections to reduce the systematic uncertainty of both the measurements and the theoretical calculations needed to extract the in-medium structure functions.

We propose to measure the semi-inclusive cross section with the scattered electron detected simultaneously at two kinematics: one corresponding to $x$'~0.3 where the EMC effect is negligible and little nucleon modification is expected and the other at $x$'~0.5 where the EMC effect is large and significant nucleon modification is expected (in some models). The scattered electrons will be detected in coincidence with recoil protons and neutrons with momentum of 300-600 MeV/c over a wide backward angular range $85° < \theta_s < 174°$.

## III.2 Experimental setup

We propose to perform the measurement in Hall C using the two high-resolution spectrometers (SHMS and HMS) in coincidence with a dedicated large-angle 1.5-sr proton and neutron detector LAD (see section III.2.2 and Appendix I). The incident beam energy is 10.9 GeV.

For the deuteron in-medium structure function measurement, the two spectrometers are set at 4.4 GeV / $13.5^0$ and 4.4 GeV / $17^0$ on both sides of the beam, to measure the electrons at the low and high $x$' kinematics simultaneously. To reduce systematic uncertainties in the measurement each one of the spectrometers will be set at both kinematics for part of the time. These kinematic settings were chosen to maximize the number of "golden events" which fulfill the following requirements:

$Q^2 \geq 2$ GeV$^2$

$W'^2 \geq 4$ GeV$^2$

$\theta_{pq} \geq 110^0$

$p_s \geq 275$ MeV/$c$ (protons)

$p_s \geq 225$ MeV/$c$ (neutrons)

### III.2.1 SHMS and HMS

The kinematic settings for the measurements are shown in Table 1 and Fig. 12.

| Low *x'* range (central values): | High *x'* range (central values): |
|---|---|
| $E_{in} = 10.9$ GeV | $E_{in} = 10.9$ GeV |
| $E' = 4.4$ GeV | $E' = 4.4$ GeV |
| $\theta_e = 13.5°$ | $\theta_e = -17°$ |
| $Q^2 = 2.65$ GeV$^2$ | $Q^2 = 4.19$ GeV$^2$ |
| $|\vec{q}| = 6.7$ GeV/$c$ | $|\vec{q}| = 6.8$ GeV/$c$ |
| $\theta_q = -8.8°$ | $\theta_q = 10.8°$ |
| $x = 0.217$ | $x = 0.34$ |

**Table 1:** Experimental Electron Kinematics. The minus signs indicate that the two spectrometers are on opposite sides of the beam line and each of the momentum transfer angles is on the opposite side of the beam line from the corresponding spectrometer.

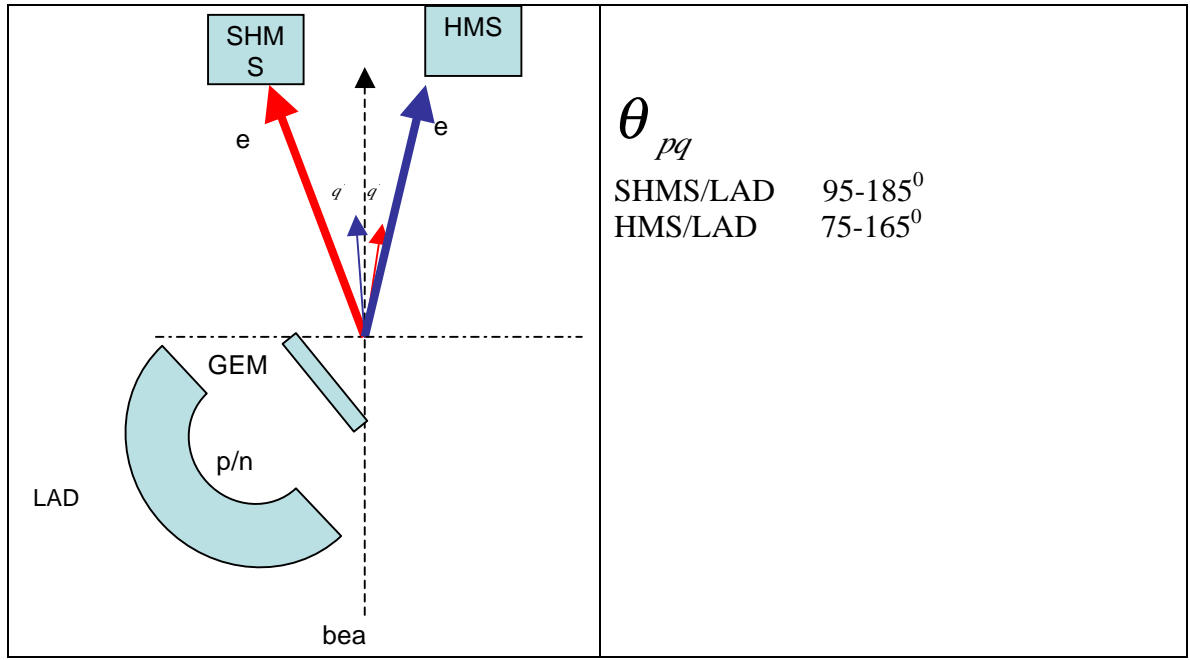

**Figure 12:** Schematic of the proposed experimental setup in Hall C. The role of the HMS and SHMS as the low and high *x'* detectors will be interchanged during the experiment. The LAD will span an angular range from about 85 to 177° in the lab. The GEM detector shown is a possibility if we need to improve our angular resolution.

The acceptance of the two electron spectrometers at the two kinematic points is shown in Fig. 13. Note that of the 34 days (820 hours) of beam, we plan to use 300 hours with the HMS at 13.5° and the SHMS at 17°, 300 hours with the HMS at 17° and the SHMS at 13.5°, and 220 hours with both spectrometers at 17°.

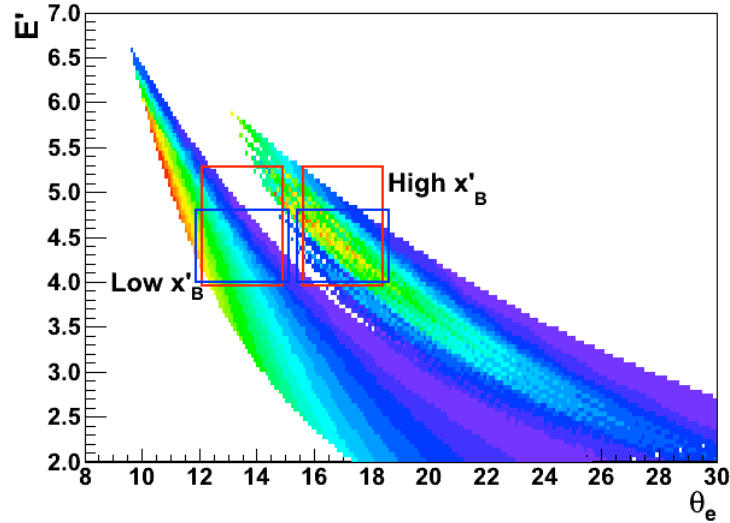

**Figure 13:** The SHMS and HMS acceptance at the selected kinematics. Events are distributed according to the cross section [40]. Red boxes show the SHMS acceptance, blue the HMS. High $x'$ means $x'>0.45$, low $x'$ means $x'<0.35$. The color indicates the relative number of events. The scales for the high and low $x'$ regions are different.

### III.2.2 LAD - the recoil nucleon detector

To detect recoil nucleons, we propose to use a 1.5-sr Large Acceptance Detector (LAD). In order to save money and effort, this detector will be assembled from the scintillator counters (TOF) of the current CLAS (CLAS6) that will not be used as part of CLAS12. We propose to use all 6 sectors of panels 3 and 4 of the TOF system, amounting to 138 scintillators.

Using these detectors for the LAD requires that they be removed intact when CLAS6 is decommissioned. A decision to save these detectors must be taken before decommissioning starts in May 2012.

For this measurement, we plan to cover scattering angles from 85° to 174° on the beam-left side at a distance of about 4 meters from the target. The approximately 4-m long scintillator bars will then give an out-of-plane angular coverage of about ±25° (except from about 85° to 112° where the SHMS structure reduces the acceptance to about ±10°). We propose to stack two thicknesses of the TOF counters at smaller angles (85° to 112°), and five thicknesses at larger angles (111° to 174°). This will give a total solid angle of about 1.5 sr and a neutron detection efficiency of about 20% at the larger angles. Note that the LAD is very flexible and its optimized final configuration will differ slightly from that presented here. We will also pursue the possibility of increasing our neutron detection efficiency by using the 30 CLAS6 TOF detectors from panel-2 that will be unused in CLAS12.

By measuring the energy deposited by protons in the first two layers of the LAD, we can perform a crude momentum measurement that will allow us to place tighter cuts on the proton time-of-flight and significantly improve our signal-to-noise.

More details about the LAD detector can be found in Appendix I.

As shown in Fig. 12 we are also considering adding a GEM detector closer to the target to improve the angular resolution of the recoil protons. See Appendix I for a discussion on the impact of the GEM on the $x'$ measurement uncertainty.

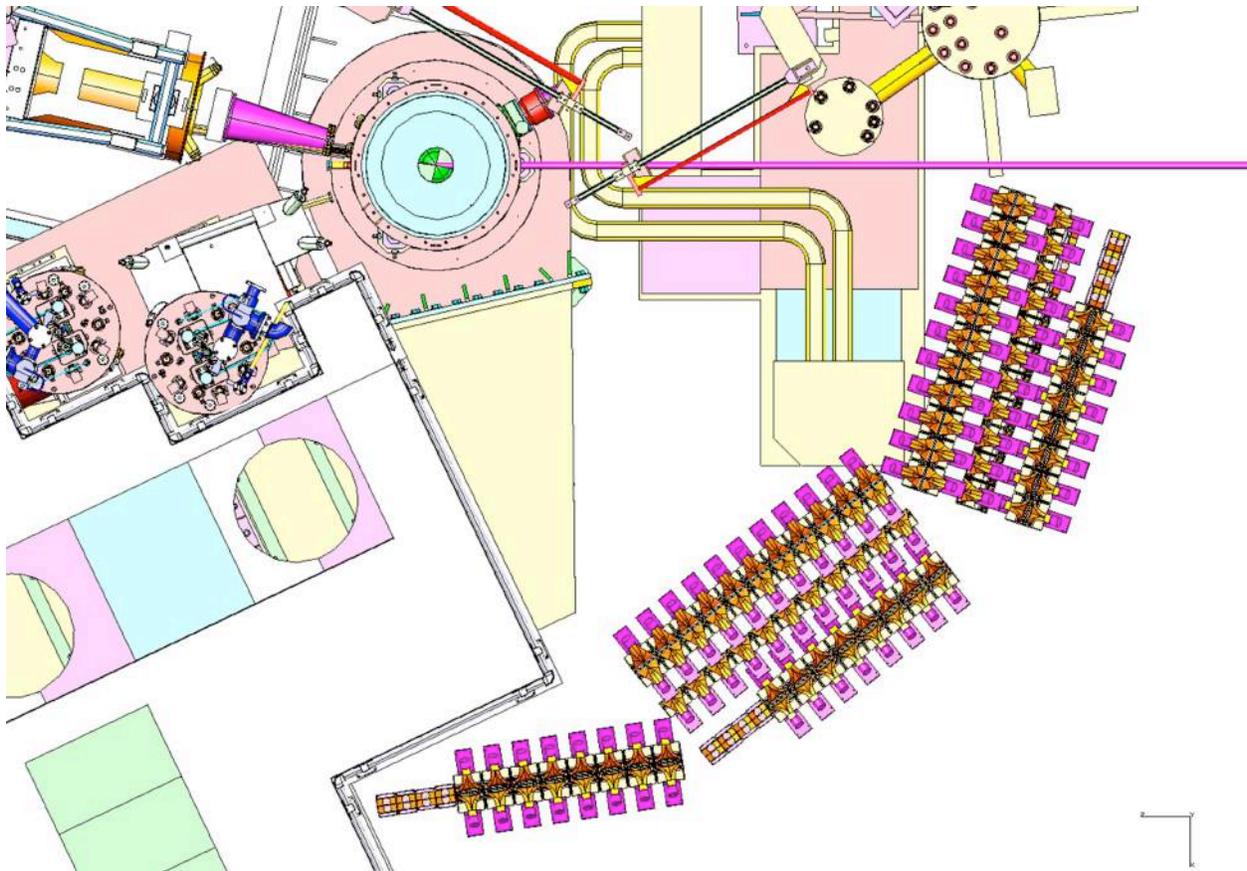

**Figure 14:** Plan view of the layout of the LAD detector in Hall C. The incoming beam line is shown by the purple horizontal line in the upper left. The scattering chamber is shown by the light blue circle in the middle toward the top of the figure. The HMS entrance is shown in the upper left and the SHMS structure is shown on the left side. There are three sectors of LAD. The first, at scattering angles from 85 to 113°, consists of two back-to-back planes of scintillators, comprised of two sectors of the current "panel 4" of the CLAS6 TOF. The middle sector, covering angles from 110 to 145°, contains a back-to-back pair of "panel-3" TOF sectors, followed by a single layer of panel-4 and a double layer of panel-3 for a total of five thicknesses of scintillator (although with diminishing solid angle as the distance to the target increases). The third sector covers angles from 145 to 175° with the same five thicknesses of scintillator as the middle sector. Layout courtesy of M. Fowler.

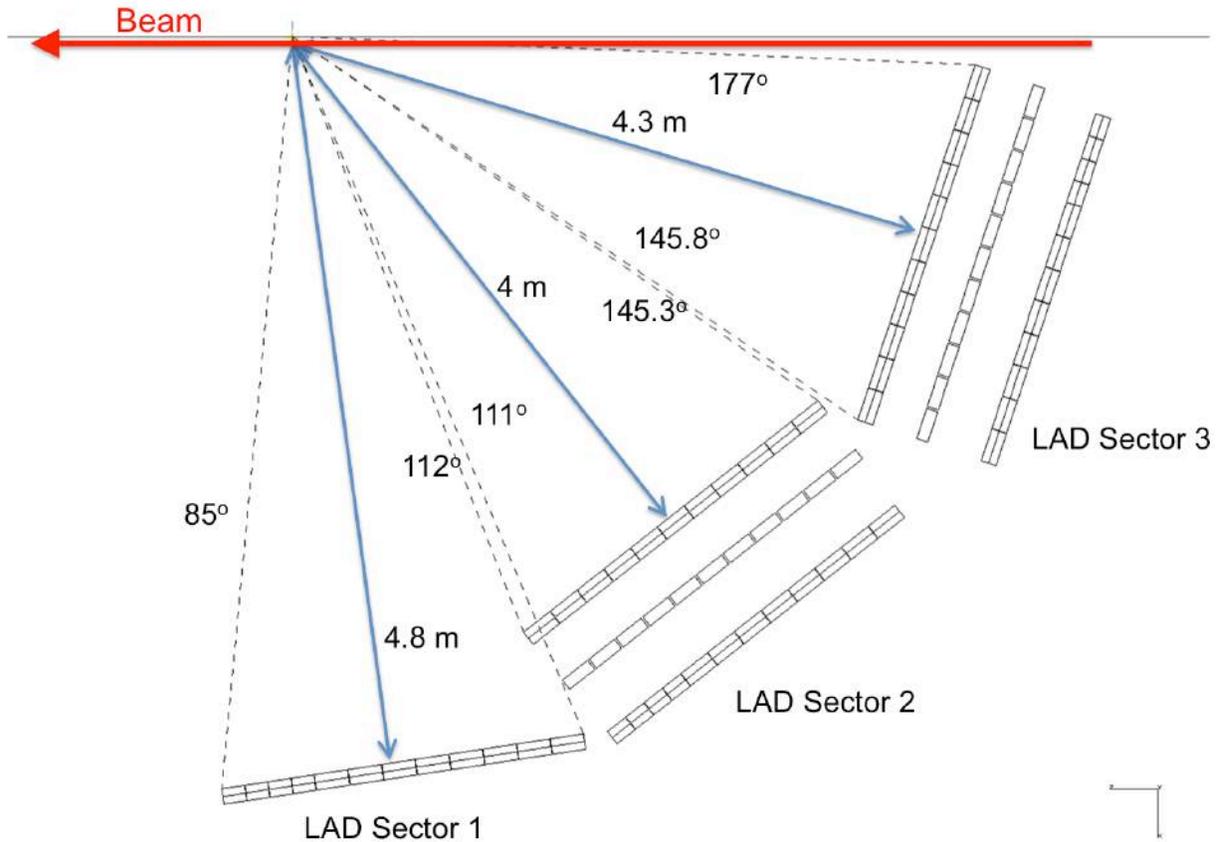

**Figure 15:** LAD angular coverage. The angles shown are with respect to the beam direction.

The layout and acceptance of the LAD are shown in Figs. 14, 15 and 16.

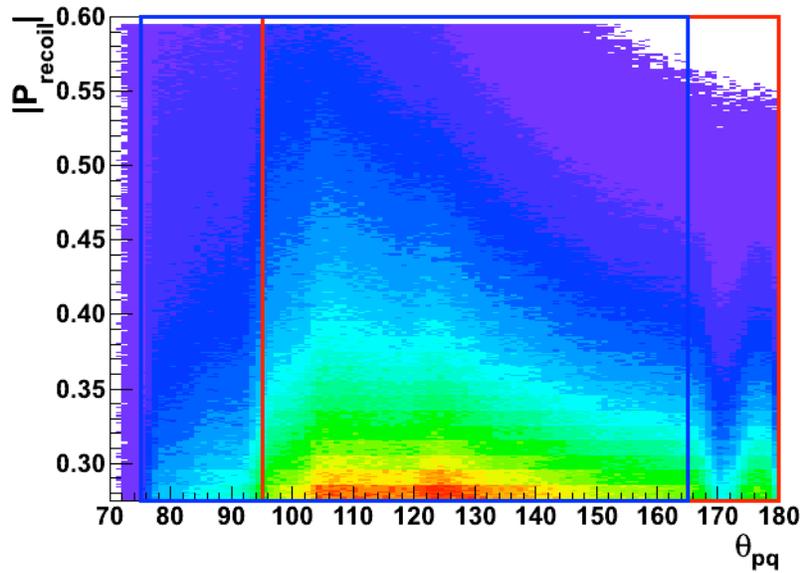

**Figure 16:** The LAD acceptance. Events are distributed according to the cross section [40]. The red lines show the $\theta_{pq}$ acceptance of the LAD in coincidence with the SHMS, the blue lines with HMS.

### III.2.3 Scattering chamber

The scattering chamber will be similar to the standard Hall C 12 GeV scattering chamber, but with windows cut out on the beam left side to allow a large out-of-plane acceptance for the LAD for $80^o < \theta_s < 174^o$, similar to the Hall A chamber used for BigBite [43]. For an out-of-plane angle acceptance of $\pm 25^o$ and the Hall C scattering chamber radius of 22.5 inches, the cut out windows would be about $2 \times 22.5" \times \tan(25^0) = 21"$ vertically. If necessary, the chamber can be strengthened by (a) adding moveable aluminum vertical support rods, (b) adding horizontal support rods, and/or (c) increasing the scattering chamber wall thickness. See more details on the scattering chamber in appendix II.

### III.2.4 Experimental Acceptance

The two spectrometers' angular and momentum coverage are shown in Fig. 13. The phase space coverage as a function of $Q^2, x', W'$ and the spectator angle $\theta_{sq}$ and momentum $p_s$ is shown in Fig 17.

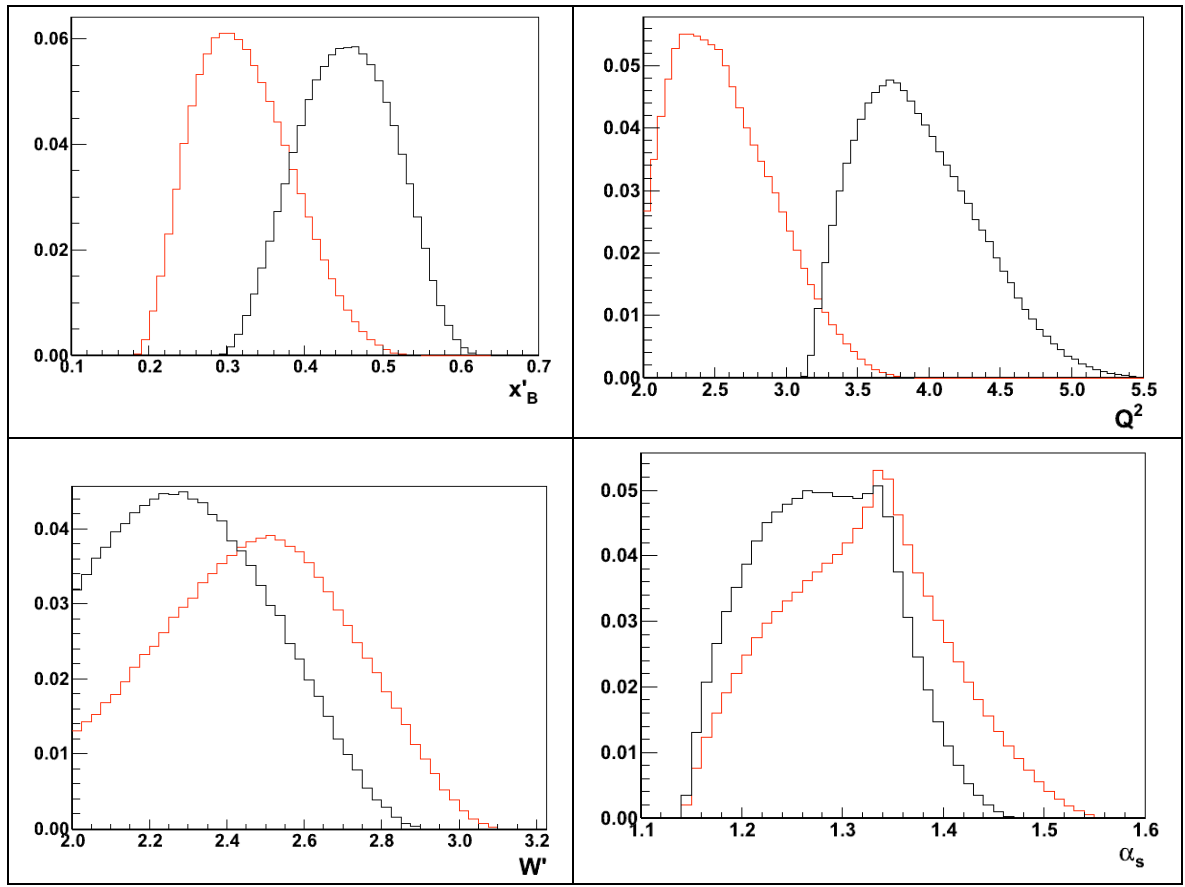

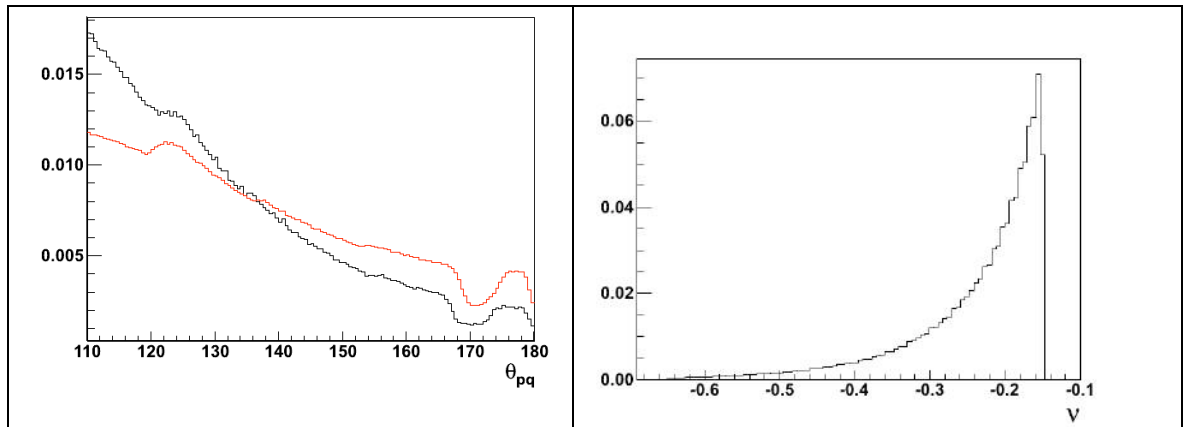

**Figure 17:** The phase space covered by the two proposed settings. The red and black curves show the low and high *x'* settings respectively. Events simulated here were detected by both HMS and SHMS. All plots are normalized to an area of 1. Note that bumps in the different distributions come from combining the HMS/LAD and SHMS/LAD acceptances.

**III.3 Extracting the in medium structure function**

As stated above, the goal of the experiment is to measure the in-medium nucleon structure function and to compare it to the free one. The obstacles are: (1) The deuteron wave function is not well known at the large spectator momenta that we plan to measure. This uncertainty increases with spectator nucleon momentum, $p_s$. (2) The FSI are not well known. These are strong functions of $W'$ and $p_s$.

However, we will measure cross sections over a wide kinematic range to provide data that will allow theorists to study the variation of the cross section with different variables, allowing them to optimize different ingredients of their calculations.

Effects from off shell nucleons should be small at low $x'$. We will fix $p_s$ and $x'$ to significantly reduce the uncertainty from the deuteron momentum distribution. Then we will vary $W'$ to study how FSI varies with the invariant mass of the produced hadrons. Figure 18 shows the $W'$ distribution for one bin in low $x'$ and small $p_s$. Reproducing this data by the theoretical calculation is a strong test of its ability to describe FSI.

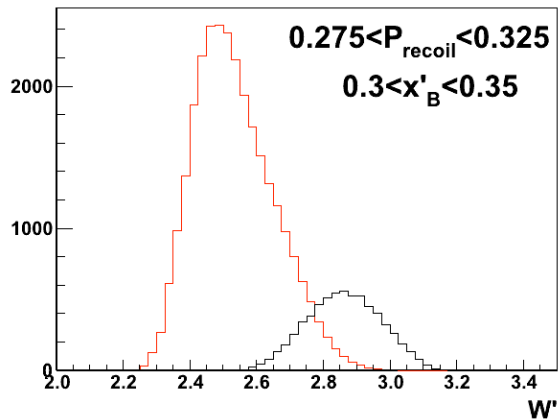

**Figure 18:** The expected $d(e,e'p_s)$ $W'$ distribution for one bin in $x'$ and recoil momentum, $p_s$. The red shows the low $x'$ and the black shows the high $x'$ setup. Data from both spectrometers was summed.

Similarly, measuring the cross section at fixed $x'$ and $W'$ and varying $p_s$ will let the theorists constrain the recoil momentum distribution used in their calculations. See Fig. 19 for the expected data.

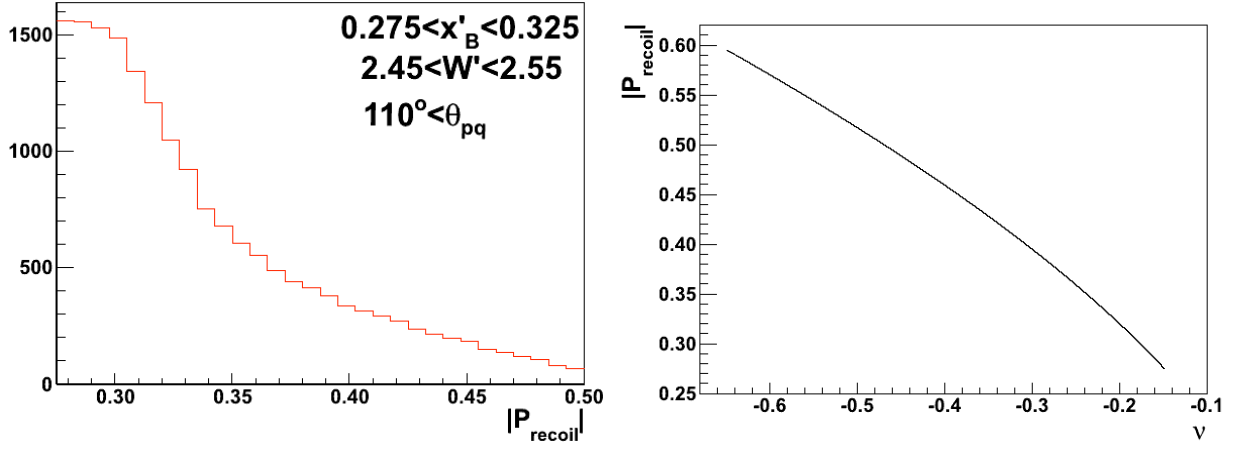

**Figure 19**: (left) The expected $d(e,e'p_s)$ recoil momentum ($p_s$) distribution for one bin in $W'$ and $x'$. (right) The relationship between virtuality and recoil nucleon momentum.

After checking the calculation at low $x'$ where nucleon off-shell effects are expected to be small, we can look for off shell effects by fixing both $W'$ and $p_s$ and measuring the ratio of cross sections at high and low $x'$. The simulated data for this are shown in Fig. 20.

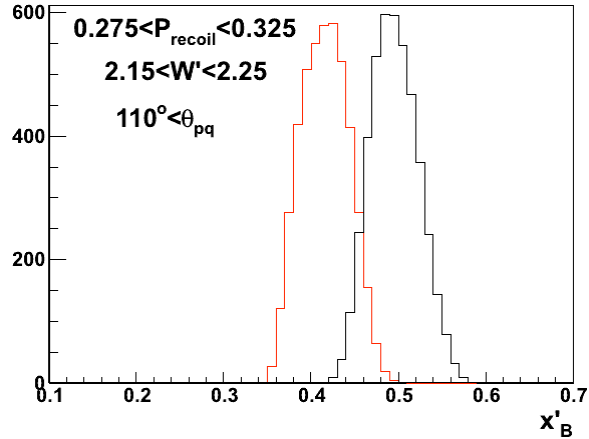

**Figure 20:** The expected $d(e,e'p_s)$ data for one bin at small $p_s$ and $W'$ for the two $x'$ settings. These conditions should minimize the uncertainty of both FSI and deuteron momentum distribution contributions, especially in the ratio of cross sections.

After optimizing the state of the art cross section calculations without nucleon modification at low $x'$, we will divide the measured $d(e,e'p_s)$ cross section ratio at large and small $x'$ by the calculated ratio to get (Equation 1):

$$(\frac{d^4\sigma}{dx_1 dQ^2 d\vec{p}_2} / \frac{d^4\sigma^{calc}}{dx_1 dQ^2 d\vec{p}_2}) \cdot (\frac{d^4\sigma}{dx_2 dQ^2 d\vec{p}_2} / \frac{d^4\sigma^{calc}}{dx_2 dQ^2 d\vec{p}_2}) =$$
$$[F_2^*(x_1',\alpha_s,p_T,Q_1^2) / F_2^*(x_1',\alpha_s,p_T,Q_1^2)_{calc}] \cdot [F_2^*(x_2',\alpha_s,p_T,Q_1^2) / F_2^*(x_2',\alpha_s,p_T,Q_1^2)_{calc}]$$

We will then plot this double ratio versus nucleon virtuality and also versus spectator light cone momentum fraction $\alpha_s$. This double ratio should be very sensitive to $F_{2N}^{eff}/F_{2N}$, the ratio of the in-medium and free nucleon structure functions

### III.4 Interpretation Uncertainties

To extract the ratio of the in medium nucleon structure functions to the free structure functions there are theoretical uncertainties related to the knowledge of the free structure functions as well as to the relation between the nucleon structure function and the deuteron semi inclusive cross section. The two main contributions to the latter are the knowledge of the high momentum tail of the deuteron wave function and the knowledge of the FSI. By measuring cross sections over the large kinematical coverage of this measurement, we will (with the help of our theory colleagues) also be able to tune and check the calculations.

The FSI uncertainty is dominant. To estimate it we use the calculations that predict the largest FSI contributions at the backward angles [41,42] and compare the ratio of cross section calculated with PWIA to the same ratio calculated with FSI. The difference between the ratio of PWIA+FSI cross sections and the ratio of PWIA cross sections at the proposed kinematics is less than 4%. Thus, the theoretical uncertainties in the ratio due to FSI should be small.

The uncertainties in the free structure functions of the proton are small. The free neutron structure functions are much less well known but will be measured by the BoNuS experiment [32,33].

## III.5 Rates

The signal and background rates were calculated by a simulation that took into account the kinematics and acceptance of the detectors. In the following, the deuteron luminosity is assumed to be $10^{36}$ cm$^{-2}$ sec$^{-1}$. The input to the simulation is presented below.

### III.5.1 Semi-Inclusive rates

We calculated the differential cross sections $d^4\sigma / dE'_e d\Omega_e dT_s d\Omega_s$ using the PWIA model of Cosyn and Sargsian [40] for $d(e,e'p_s)$ and $d(e,e'n_s)$ for the proposed kinematics, in units of nb/sr$^2$-GeV$^2$. Note that the Cosyn/Sargsian PWIA cross section agrees closely with that of Ciofi degli Atti and collaborators [41]. Note also that the PWIA model should underpredict the cross section at angles with large Final State Interaction contributions ($\theta_{ps} \approx 90^o$).

The expected uncertainties for the ratio of the in-medium to free nucleon response functions as calculated from the simulated data according to Eq. 1 are shown in Fig. 21.

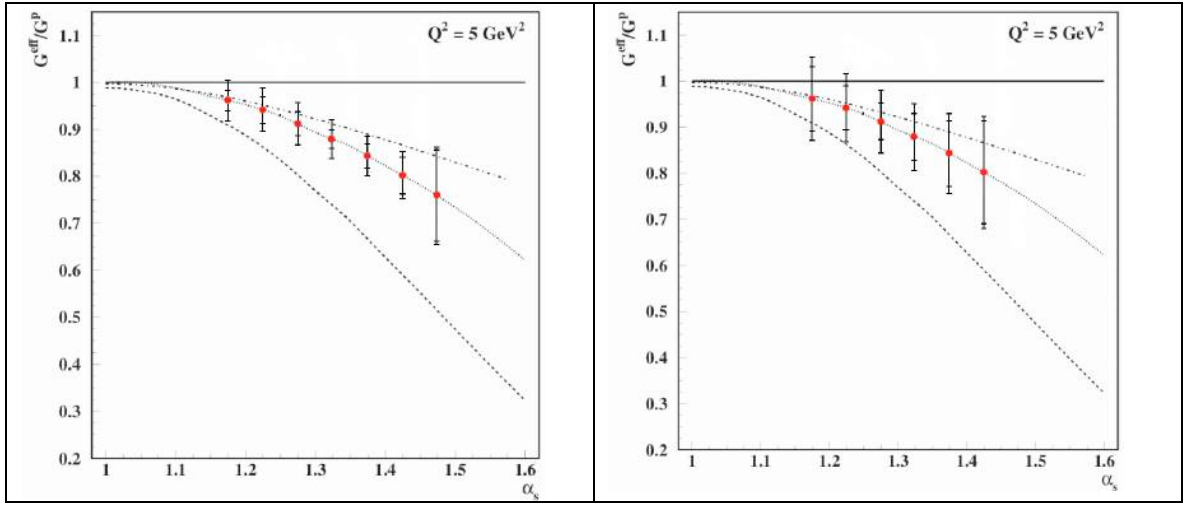

**Figure 21:** Model predictions of the ratio $F_{2N}^{eff} / F_{2N}$ as in Fig. 8 with simulated data including statistical (inner error bars) and systematical (outer error bars) uncertainties for 34 days of data. The left panel is for the $d(e,e'p_s)$ reaction and the right for $d(e,e'n_s)$. Note that we expect an additional 4% interpretation uncertainty due to uncertainty of the relative effects of FSI in the measured ratios. Note also that the "$Q^2$ = 5 GeV$^2$" refers to the models, not the data.

Tables 2 to 7 show the number of semi-inclusive $d(e,e'p_s)$ and $d(e,e'n_s)$ events expected in 34 days hours of beam (300 hours of each spectrometer at low x' and 520 hours at high x'), as a function of $\alpha_s$ (Tables 2 and 5) and as a function of angle and momentum, $\theta_{sq}$ and $p_s$ (Tables 3, 4, 6 and 7). The statistical uncertainties are also shown. The events include those detected by both the SHMS and the HMS. The nuclear luminosity assumed for these estimates is $10^{36}$ cm$^{-2}$ sec$^{-1}$.

| x'$_B$ \ alpha_s | 1.15-1.2 | 1.2-1.25 | 1.25-1.3 | 1.3-1.35 | 1.35-1.4 | 1.4-1.45 | 1.45-1.5 |
|---|---|---|---|---|---|---|---|
| x'$_B$>0.45 | 2230 (4.3%) | 5445 (2.9%) | 8290 (2.5%) | 12160 (2.1%) | 8590 (2.7%) | 3030 (4.6%) | 355 (12.6%) |
|  | 7675 (2.1%) | | 20450 (1.6%) | | 11620 (2.3%) | | |
|  | | | | | | | |
| 0.25<x'$_B$<0.35 | 14025 (1.3%) | 27330 (0.97%) | 32590 (0.93%) | 41905 (0.8%) | 29115 (1%) | 12700 (1.5%) | 4160 (2.6%) |
|  | 41355 (0.6%) | | 74495 (0.6%) | | 41815 (0.8%) | | |

**Table 2:** Expected number of events and associated statistical uncertainties for $d(e,e'p_s)$ for $Q^2 > 2$ GeV$^2$, $W' > 2$ GeV and $\theta_{ps} > 110°$.

| $\theta_{pq} \setminus P_{recoil}$ | 275-325 | 325-375 | 375-425 | 425-475 | 475-525 | **Total** |
|---|---|---|---|---|---|---|
| 80-90 | 95 (18.8%) | 80 (22%) | 80 (22.5%) | 90 (20.2%) | 100 (19%) | 445 (9.1%) |
| 90-100 | 435 (9%) | 370 (10.5%) | 360 (11.3%) | 375 (10.4%) | 395 (10%) | 1995 (4.5%) |
| 100-110 | 1585 (5.2%) | 1290 (6.1%) | 1170 (6.6%) | 1100 (6.5%) | 970 (7%) | 6115 (2.8%) |
| **110-120** | 2475 (4.2%) | 1935 (5%) | 1600 (5.8%) | 1250 (6.5%) | 770 (7.2%) | **8030** (2.5%) |
| **120-130** | 3250 (3.8%) | 2350 (4.8%) | 1644 (6.2%) | 870 (8.1%) | 210 (15.4%) | **8325** (2.5%) |
| **130-140** | 2940 (4.4%) | 1715 (6.6%) | 815 (10.4%) | 175 (20.3%) | N/A | **5645** (3.5%) |
| **140-150** | 4055 (3.3%) | 2040 (5.2%) | 570 (10.2%) | 20 (48.6%) | N/A | **6685** (2.7%) |
| **150-160** | 3750 (3.5%) | 1470 (6.2%) | 180 (17.7%) | N/A | N/A | **5400** (3%) |
| **160-170** | 2510 (4.7%) | 830 (8.6%) | 45 (34.7%) | N/A | N/A | **3385** (4.1%) |
| **170-180** | 1850 (6.4%) | 395 (14.9%) | N/A | N/A | N/A | **2245** (5.9%) |
| **Total $\theta_{pq} > 110°$** | **20830** (1.6%) | **10735** (2.3%) | **4855** (3.6%) | **2315** (4.9%) | **980** (7.3%) | **39715** (1.2%) |
| **Total** | 22945 (1.5%) | 12475 (2.1%) | 6465 (3%) | 3880 (3.6%) | 2445 (4.4%) | 48210 (1%) |

**Table 3:** The number of semi-inclusive $d(e,e'p_s)$ events as a function of the recoil angle [deg] and momentum [MeV/c], for $x'>0.45$ for 34 days of data taking at a luminosity of $10^{36}$ cm$^{-2}$ s$^{-1}$. The expected statistical uncertainties including background events are shown in parentheses. Event totals for angles greater than 110° (where we expect minimal FSI) are highlighted.

| $\theta_{pq}$ \ $P_{recoil}$ | 275-325 | 325-375 | 375-425 | 425-475 | 475-525 | Total |
|---|---|---|---|---|---|---|
| 70-80 | 2525 (2.8%) | 1600 (3.7%) | 1180 (4.4%) | 940 (4.8%) | 770 (5.2%) | 7015 (1.7%) |
| 80-90 | 5850 (1.8%) | 3680 (2.4%) | 2670 (3%) | 2125 (3.2%) | 1740 (3.5%) | 16065 (1.1%) |
| 90-100 | 8625 (1.6%) | 5480 (2.1%) | 4055 (2.5%) | 3280 (2.7%) | 2735 (2.9%) | 24175 (1%) |
| 100-110 | 14435 (1.3%) | 9160 (1.7%) | 6755 (2%) | 5350 (2.2%) | 4295 (2.4%) | 39995 (0.8%) |
| **110-120** | 15255 (1.2%) | 9590 (1.7%) | 6790 (2%) | 4995 (2.3%) | 3410 (2.7%) | **40040** (0.8%) |
| **120-130** | 16275 (1.2%) | 9725 (1.7%) | 6220 (2.2%) | 3665 (2.7%) | 1675 (4%) | **37560** (0.8%) |
| **130-140** | 10685 (1.7%) | 5810 (2.4%) | 3020 (3.5%) | 1225 (5.2%) | 290 (10.3%) | **21030** (1.2%) |
| **140-150** | 13620 (1.3%) | 6290 (2.1%) | 2315 (3.6%) | 485 (7.8%) | N/A | **22710** (1.1%) |
| **150-160** | 12650 (1.4%) | 4950 (2.3%) | 1325 (4.7%) | 100 (16.1%) | N/A | **19025** (1.1%) |
| **160-170** | 9385 (1.6%) | 3230 (2.9%) | 645 (6.8%) | N/A | N/A | **13260** (1.4%) |
| **170-180** | 4115 (1.9%) | 1125 (6.1%) | 160 (16.8%) | N/A | N/A | **1285** (2.6%) |
| **Total $\theta_{pq} > 110°$** | **77870** (0.56%) | **40720** (0.84%) | **20475** (1.2%) | **10470** (1.6%) | **5375** (2.2%) | **154910** (0.41%) |
| **Total** | 109305 (0.46%) | 60640 (0.67%) | 35135 (0.9%) | 22165 (1.1%) | 14910 (1.3%) | 242160 (0.32%) |

**Table 4:** Same as above for $d(e,e'p_s)$ and 0.25< $x$'<0.35.

| x'$_B$ \ alpha_s | 1.15-1.2 | 1.2-1.25 | 1.25-1.3 | 1.3-1.35 | 1.35-1.4 | 1.4-1.45 | 1.45-1.5 |
|---|---|---|---|---|---|---|---|
| x'$_B$>0.45 | 1470 (7%) | 3390 (4.8%) | 4825 (4%) | 3260 (5.4%) | 1760 (7.8%) | 600 (13.2%) | 65 (34.1%) |
| | 4860 (4%) | | 8085 (3.3%) | | 2360 (6.7%) | | |
| 0.25<x'$_B$<0.35 | 8090 (2.4%) | 12860 (1.9%) | 14380 (1.8%) | 8425 (2.5%) | 4450 (3.5%) | 2005 (5.2%) | 685 (8.6%) |
| | 20950 (1.5%) | | 22805 (1.5%) | | 6455 (2.9%) | | |

**Table 5:** Expected number of events and associated statistical uncertainties for $d(e,e'n_s)$ for $Q^2 > 2$ GeV$^2$, $W' > 2$ GeV and $\theta_{ps} > 110^o$.

| $\theta_{pq}$ \ P$_{recoil}$ | 225-275 | 275-325 | 325-375 | 375-425 | 425-475 | 475-525 | **Total** |
|---|---|---|---|---|---|---|---|
| 100-110 | 320 (14.8%) | 195 (22.8%) | 140 (30.2%) | 135 (30.6%) | 135 (29.7%) | 120 (31.7%) | 1045 (9.9%) |
| **110-120** | 490 (12.1%) | 325 (16.9%) | 225 (23%) | 190 (25.8%) | 155 (28.8%) | 95 (37.1%) | **1480** (8.2%) |
| **120-130** | 1090 (8.2%) | 735 (11%) | 525 (14%) | 370 (17.4%) | 190 (24.2%) | 45 (46.9%) | **1865** (5.5%) |
| **130-140** | 1510 (7%) | 930 (9.8%) | 570 (13.7%) | 285 (19.8%) | 60 (41.3%) | N/A | **3355** (5.1%) |
| **140-150** | 1425 (7.2%) | 800 (10.9%) | 410 (16.4%) | 110 (31.3%) | N/A | N/A | **2745** (5.6%) |
| **150-160** | 1275 (7.7%) | 660 (12.2%) | 255 (21.1%) | 30 (58.3%) | N/A | N/A | **2220** (6.3%) |
| **160-170** | 860 (9.5%) | 430 (15.4%) | 135 (28.5%) | N/A | N/A | N/A | **1425** (7.9%) |
| **170-180** | 905 (7.6%) | 395 (13.3%) | 85 (30.3%) | N/A | N/A | N/A | **1385** (6.5%) |
| **Total** $\theta_{pq}>110^o$ | **6465** (3.1%) | **4275** (4.6%) | **2205** (7%) | **985** (10.7%) | **405** (17%) | **140** (29.4%) | **14475** (2.36%) |
| **Total** | 6785 (3%) | 4470 (4.5%) | 2345 (6.8%) | 1120 (10.1%) | 540 (14.7%) | 260 (21.5%) | 15520 (2.3%) |

**Table 6:** Same as Table 3 for spectator neutrons $d(e,e'n_s)$ and x'>0.45.

| $\theta_{pq}$ \ $P_{recoil}$ | 225-275 | 275-325 | 325-375 | 375-425 | 425-475 | 475-525 | Total |
|---|---|---|---|---|---|---|---|
| 70-80 | 210 (13.6%) | 125 (18.5%) | 75 (25.7%) | 35 (45.7%) | N/A | N/A | 445 (10.2%) |
| 80-90 | 445 (9.4%) | 255 (13.3%) | 165 (17.2%) | 105 (22.9%) | 60 (34.4%) | 45 (40.5%) | 1075 (6.6%) |
| 90-100 | 765 (7.3%) | 435 (10.4%) | 295 (13%) | 220 (15.1%) | 175 (16.8%) | 145 (18.2%) | 2035 (4.7%) |
| 100-110 | 2810 (3.9%) | 1525 (5.8%) | 980 (7.6%) | 720 (8.9%) | 575 (9.8%) | 460 (10.7%) | 7070 (2.7%) |
| **110-120** | 3245 (3.7%) | 1765 (5.5%) | 1140 (7.1%) | 835 (8.3%) | 625 (9.5%) | 445 (11%) | **8055** (2.5%) |
| **120-130** | 5665 (2.8%) | 3030 (4.2%) | 1830 (5.7%) | 1180 (7%) | 700 (9%) | 335 (12.6%) | **12740** (2%) |
| **130-140** | 5450 (2.9%) | 2740 (4.4%) | 1490 (6.3%) | 775 (8.7%) | 315 (13.4%) | 75 (26.1%) | **10845** (2.2%) |
| **140-150** | 4410 (3.2%) | 2075 (5.1%) | 975 (7.7%) | 370 (12.7%) | 85 (25.6%) | N/A | **7915** (2.5%) |
| **150-160** | 3565 (3.6%) | 1550 (5.9%) | 610 (9.8%) | 170 (18.6%) | N/A | N/A | **5895** (2.9%) |
| **160-170** | 2595 (4.2%) | 1085 (7.1%) | 370 (12.7%) | 75 (28%) | N/A | N/A | **4125** (3.5%) |
| **170-180** | 1915 (4.3%) | 675 (8%) | 190 (15.9%) | 30 (39.6%) | N/A | N/A | **2810** (3.7%) |
| **Total $\theta_{pq} > 110°$** | **26845** (1.3%) | **12920** (2%) | **6605** (3%) | **3435** (4.1%) | **1725** (5.7%) | **855** (7.9%) | **52385** (0.98%) |
| Total | 31075 (1.2%) | 15260 (1.9%) | 8120 (2.7%) | 4515 (3.6%) | 2535 (4.7%) | 1505 (6%) | 63010 (0.9%) |

**Table 7:** Same as above for spectator neutrons $d(e,e'n_s)$ and $0.25 < x' < 0.35$.

### III.5.2 Protons Singles Rates

Since the backward protons are primarily spectators to the reaction, the backward proton rate should not increase significantly with the electron beam energy (since the Mott cross section decreases as $E^{-2}$).

The singles proton rates are estimated in two ways based on our experience with the SRC (E01-015) experiment. We measured the backward angle protons in a dedicated test run at $\theta_p \approx 100^o$ and we measured the proton rates in the BigBite spectrometer at $\theta_p \approx 90^o$ during the experiment.

In a test using the HRS at $100^o$ [46], we found proton singles rates of $10^5$ Hz for momentum $p >$ 0.25 GeV/c, nucleon luminosity of $3.8 \times 10^{37}$ cm$^{-2}$ s$^{-1}$ (100 µA beam incident on a 0.5 mm C target) and a solid angle of 6 msr. We will conservatively assume that the proton single-arm cross section is constant with scattering angle for $\theta_p \geq 100^o$.

For the 1.5-sr LAD, the expected rate is:

$$R(\text{singles protons}) = \frac{(10^5 \text{Hz})(2 \times 10^{36} \text{cm}^{-2}\text{s}^{-1})}{(3.8 \times 10^{37} \text{cm}^{-2}\text{s}^{-1})(6 \times 10^{-3} \text{sr})} = 9 \times 10^5 \text{ Hz/sr}$$

The singles rates on the $E$ counters in BigBite during E01-015 were about 80 KHz per counter at a luminosity of $L = 5 \times 10^{37}$ cm$^{-2}$s$^{-1}$ giving a rate for the LAD of:

$$R(\text{singles protons}, p_p > 0.25 \text{ GeV}/c) = \frac{(8 \cdot 10^4 \text{p/s})(2 \cdot 10^{36} \text{cm}^{-2}\text{s}^{-1})}{(5 \times 10^{37} \text{cm}^{-2}\text{s}^{-1})(4 \times 10^{-3} \text{sr})} = 8 \times 10^5 \text{ Hz/sr}$$

These two measurements are consistent, and we will therefore assume a rate of $10^6$ Hz/sr of protons above 250 MeV/c in the LAD.

Using the measured momentum distribution [46], the expected singles rates for the LAD are shown in Table 8.

| Momentum (GeV/c) | Rate (MHz/sr) |
|---|---|
| 0.275-0.325 | 0.22 |
| 0.325-0.375 | 0.17 |
| 0.375-0.425 | 0.14 |
| 0.425-0.475 | 0.10 |
| 0.475-0.525 | 0.08 |

**Table 8:** Proton singles rates anticipated in the 1.5-sr LAD.

Note that these rates should be a significant overestimate, since we used the proton cross section at $\theta_p = 100^o$ and applied it to all larger proton scattering angles.

### III.5.3 Singles inclusive (e,e') rates

We used inclusive cross sections from Ref. [40] to estimate the single electron rates. At $E_{\text{beam}}$=10.9 GeV, $E'$=4.4 GeV, and $\theta = 13.5$ and $17^o$, the cross sections are

$$\frac{d^2\sigma}{d\Omega_e dE'} = 14.1 \frac{\text{nb}}{\text{sr-GeV}}$$

and 6.8 nb/sr-GeV respectively. This gives electron singles rates, for the 5 msr angular and 30% momentum acceptance of the SHMS, of

$$R = (14 \times 10^{-33} \text{cm}^2/\text{sr-GeV})(10^{36} \text{cm}^{-2}\text{s}^{-1})(5 \times 10^{-3} \text{sr})(0.3 \times 4.4 \text{ GeV}) = 100 \text{ Hz}$$

and 50 Hz for the 13.5° cross section. The HMS rates will be comparable.

### III.5.4 Singles Neutrons

The singles neutron rate was obtained by scaling the calculations done by Pavel Degtiarenko (see Table 1 in Ref [44]). The calculations assume a detection threshold of 5 MeVee. The scaled rate for this proposal is 5 $10^5$ Hz/sr.

### III.5.5 Accidental Coincidence Rates

The accidental coincidence rate $A = R_e R_N \tau$ depends on the individual electron and nucleon singles rates and the resolving time. Since we will measure nucleon momentum using time-of-flight, our resolving time will equal the longest nucleon flight time. If the LAD is placed 4 m from the target and we are sensitive to 0.2 GeV/c nucleons, then our resolving time is

$$\tau = \frac{d}{c\beta} = \frac{d}{cp/E} = \frac{4 \text{ m}}{(3 \times 10^8 \text{ m/s})0.2} = 60 \text{ ns}$$

so that our accidental rate for neutrons is
$A_n = (100 \text{ Hz})(5 \times 10^5 \text{ Hz})(60 \text{ ns}) = 3 \text{ Hz}$
and our raw proton accidental coincidence rate is five times smaller: $A_p = 0.6$ Hz.

Note that we will use the proton energy deposition in the LAD as a crude measure of proton momentum (and hence velocity) to reduce our proton time window from 60 ns to 3 ns. (See Appendix I for details of the LAD energy and time resolution.) This will eliminate well over 90% of the proton accidental coincidences and conservatively reduce our proton accidental coincidence rate to
$A_p \approx 3 \times 10^{-2}$ Hz

Note also that these are the accidental rates over 1-sr of the LAD for all kinematics. The effect of accidental coincidences are included in the experimental simulation and in the statistical uncertainties bin-by-bin.

We will also consider the effects of using pulsed beam, one beam pulse every 64 ns, to see if that improves the signal to background ratio, especially for neutrons.

### III.6 Uncertainties

### III.6.1 Statistical errors

The statistical uncertainties shown in Tables 2-7 include contributions from the signal and the random coincidence background. The statistical errors in the tables were estimated as $\sqrt{S+B}$ where $S$ and $B$ are the number of signal and background events in each bin. This is reasonable since the random background at backward angles should be a smooth function and will be oversampled and fit.

### III.6.2 Systematic errors

We will measure both absolute cross sections and ratios of cross sections. The absolute cross sections will primarily be used to understand systematic effects in both the detectors and the calculations. We will measure the inclusive cross sections at both values of $x'$ in both the HMS and the SHMS. By comparing inclusive $d(e,e')$ cross sections measured simultaneously in the two spectrometers (and reversing the spectrometer positions) we will be able to control for the effects of the spectrometer acceptances. We will also compare these measurements to the well-known inclusive $d(e,e')$ cross sections. The systematic uncertainty for spectrometer acceptance effects is estimated as 1-2%. Note that the total luminosity as well as much of the spectrometer acceptance effects will cancel in the ratio of the cross sections at different $x'$.

In addition to the luminosity and spectrometer acceptance uncertainties, a third class of systematic uncertainty comes from the unknown acceptance and efficiency of the LAD. We plan to model the detector and to calibrate it. We will calibrate the proton detection efficiency before the experiment using cosmic rays penetrating multiple layers of scintillator and, if necessary, during the experiment using $h(e,e'p)$. We will calibrate the neutron detection efficiency using the over-determined kinematically-complete $d(e,e'np)$ reaction where the electron and proton are detected by the HMS and SHMS and the recoil neutron by the LAD. We estimate about 3% uncertainty for proton detection and 5% for the neutrons.

In summary, we estimate total 4% and 6% errors in the measurement of the cross section ratios at low and high $x'$ for the proton and the neutron, respectively. We estimate a further contribution of 4% in the correction of this ratio for Final State Interactions in order to extract the proton and neutron ratios of $F_2^{eff}/F_2$, the medium modified to free structure functions. This interpretation uncertainty is discussed in section III.4 above. For the cross section measurements (not ratios) we conservatively estimate the total statistical plus systematic uncertainty to be about 10%.

### III.7 Measurement plan and Beam time request

We request 6 days for setup and calibration and 34 days of data taking on deuterium for a total of 40 days. The experiment requires the two standard spectrometers planned for the 12 GeV Hall C and a dedicated large acceptance detector (LAD) to be constructed for this measurement and maybe for others in the future. To take advantage of the large solid angle of LAD a dedicated scattering chamber with a large backward thin window, similar to the one designed for BigBite in Hall A, is also needed.

### III.8 Collaboration

The experimental group consists of people that were actively involved in exclusive measurements of high momentum transfer reaction in coincidence with recoil particles during the 6 GeV JLab program, either with the triple coincidence SRC measurements in Hall A (E01-015, and E07-006) or with the BONUS and DEEPS measurements in CLAS (E03-012, E94-102). This proposed experiment follows the same principle as the successfully completed 6 GeV experiments mentioned above and the collaboration has the needed expertise to perform the new experiment if approved.

In order to determine any possible changes in the nucleon structure function we need to collaborate extensively with theoreticians. The two theoretical groups that have contributed significantly to this proposal are C. Ciofi degli Atti and L. Kaptari as well as W. Cosyn and M. Sargsian. They anticipate participating in the interpretation of the data. Many other theoretical groups have expressed interest in the data and its interpretation. They are listed under the heading of "Theoretical Support" on the title page of this proposal.

We would like to thank the many people who proposed letters of intent and proposals to investigate this or similar physics. The proposals and LOIs include,

1. LOI12-07-012, Tagged Neutron Structure Function in Deuterium, S. Bueltmann, S. Kuhn and K. Griffioen
2. LOI 05-014, N. Liyanage and B. Wojtsekhowski
3. LOI to PAC37, Semi-inclusive Deep Inelastic Scattering from Light Nuclei by Tagging Low Momentum Spectators, X. Zhan.

# Appendix I: The LAD configuration and performances

## I.1 Introduction

To perform the above measurements, we propose to incorporate a "backward" Large Acceptance Detector (LAD) into the 12 GeV experimental Hall C setup. This detector will measure the recoil protons in coincidence with scattered electrons detected in one of the spectrometers.

To achieve the goal with a low cost, we propose to use the TOF counters of the current CLAS that will not be used as part of CLAS12.

Figure I.1 shows the CLAS detector with the TOF counters. We propose to use the two backward panels of each sector with a total of 6 panels and 23 counters per panel. The counters are 3.5-4.5 meters long with a PMT on each end (see Table I.1). The performance of these counters is well known [45]. Some of the counters will need to be refurbished and some of the PMTs will need to be replaced. In addition, space must be found to store and refurbish the detectors between the time they are removed from Hall B and the time they are installed in Hall C.

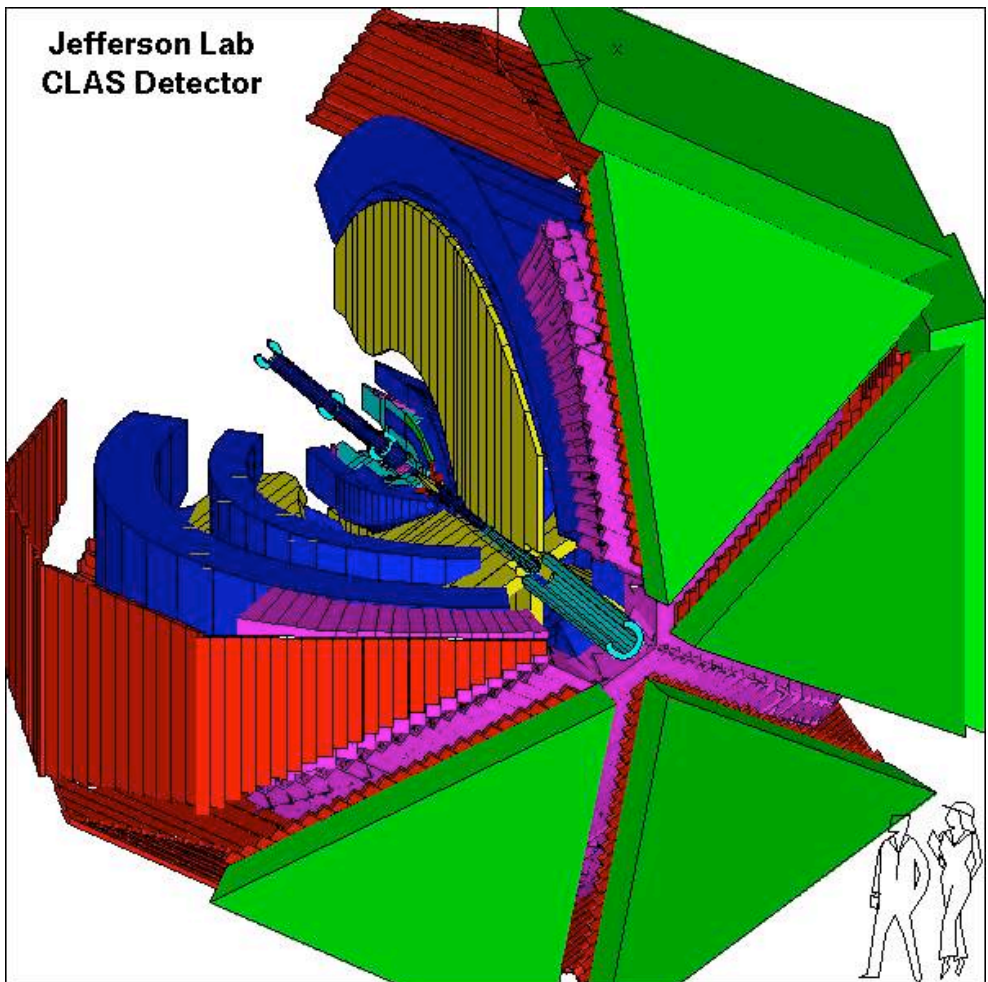

Figure I.1: The CLAS detector. The TOF panels are shown in red.

Based on available data from Hall B and simulation, we discuss some of the expected performance of the proposed LAD. The characteristics of the TOF scintillators are given in Table I.1.

Table I.1: Dimensions of the large-angle TOF scintillators. All counters are 5.08 cm thick and have a PMT on each end. "Panel-3" consists of scintillator numbers 35 to 42b and "panel-4" consists of scintillator numbers 43a to 48b.

| Scintillator number | Width (cm) | Length (cm) | Light guide configuration | PMT | Nominal lab angle (deg) | Nominal distance from target (cm) |
|---|---|---|---|---|---|---|
| 40a | 22 | 416.3 | Bent | XP4312B/D1 | 89.3 | 414 |
| 40b | 22 | 410.5 | Bent | XP4312B/D1 | 92.2 | 409 |
| 41a | 22 | 404.8 | Bent | XP4312B/D1 | 95.3 | 405 |
| 41b | 22 | 399.0 | Bent | XP4312B/D1 | 98.4 | 402 |
| 42a | 22 | 393.3 | Bent | XP4312B/D1 | 101.6 | 400 |
| 42b | 22 | 387.5 | Bent | XP4312B/D1 | 104.8 | 399 |
| 43a | 22 | 380.1 | Bent | XP4312B/D1 | 108.8 | 402 |
| 43b | 22 | 363.5 | Bent | XP4312B/D1 | 112.0 | 395 |
| 44a | 22 | 347.0 | Bent | XP4312B/D1 | 114.9 | 389 |
| 44b | 22 | 330.4 | Bent | XP4312B/D1 | 118.1 | 384 |
| 45a | 22 | 313.9 | Bent | XP4312B/D1 | 121.4 | 381 |
| 45b | 22 | 297.3 | Bent | XP4312B/D1 | 124.7 | 378 |
| 46a | 22 | 280.8 | Bent | XP4312B/D1 | 128.0 | 377 |
| 46b | 22 | 264.2 | Bent | XP4312B/D1 | 131.4 | 378 |
| 47a | 15 | 246.8 | Straight | XP2262 | 134.2 | 379 |
| 47b | 15 | 235.4 | Straight | XP2262 | 136.5 | 380 |
| 48a | 15 | 224.0 | Straight | XP2262 | 138.8 | 383 |
| 48b | 15 | 212.7 | Straight | XP2262 | 141.0 | 385 |
| 35 | 22 | 445.1 | Bent | XP4312B/D1 | 75.7 | 457 |
| 36 | 22 | 439.3 | Bent | XP4312B/D1 | 78.2 | 445 |
| 37 | 22 | 433.6 | Bent | XP4312B/D1 | 80.8 | 457 |
| 38 | 22 | 427.8 | Bent | XP4312B/D1 | 83.5 | 428 |
| 39 | 22 | 422.0 | Bent | XP4312B/D1 | 86.3 | 421 |

## I.2 Angular coverage

In general, the LAD will be designed to be flexible and to fit both Hall A and Hall C for configurations that fit the need of each particular experiment. For this measurement, we set the panels to cover the largest possible fraction of the backward hemisphere.

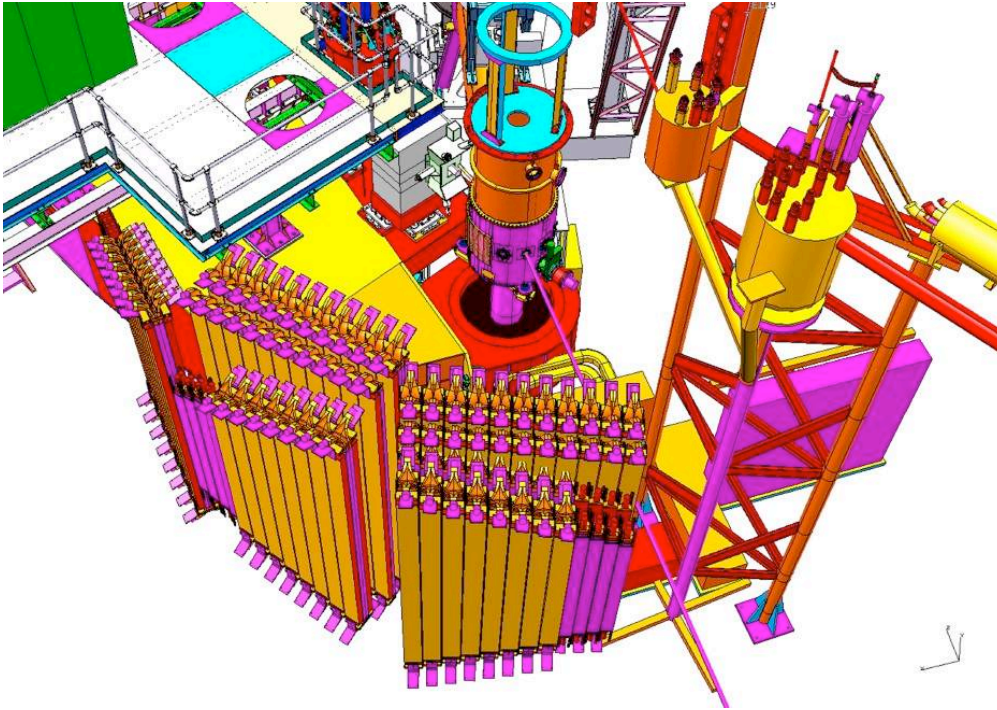

**Figure 22:** A view of the LAD from upstream. The panels of the LAD are shown in the lower left in yellow with magenta support structures. The beam line is the thin magenta line entering from the bottom, the scattering chamber is the column in the middle of the figure, parts of the SHMS structure can be seen in the upper right, and the cryogen service infrastructure is in the lower left.

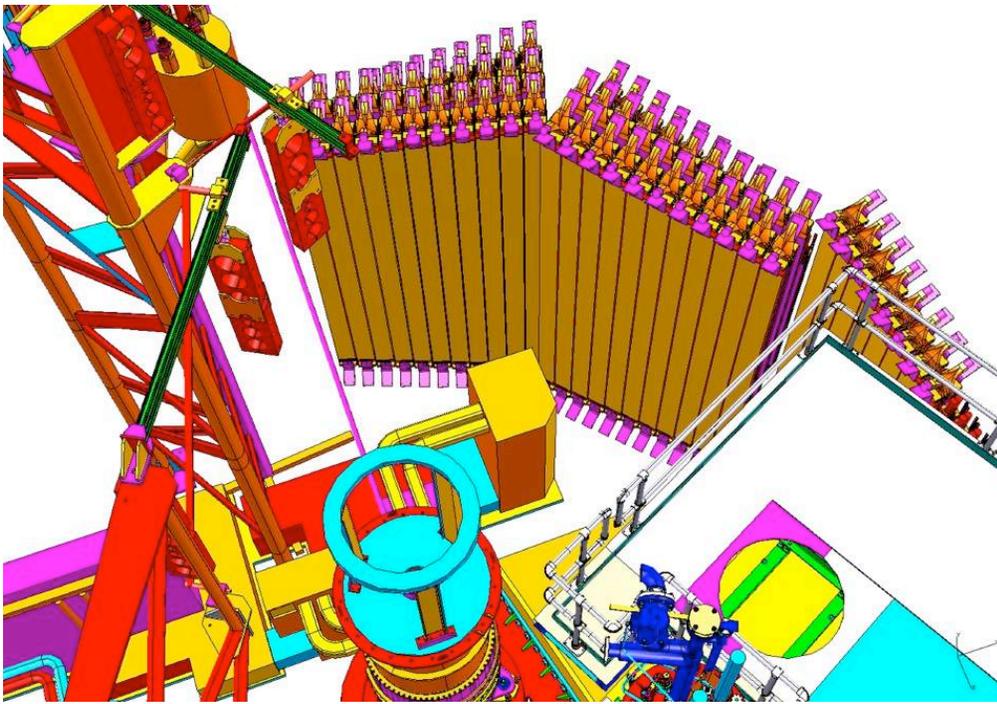

**Figure 23:** A view of the LAD from the SHMS. The top of the scattering chamber is at the bottom-middle of the figure and the SHMS platform is in the lower-right. The LAD can be seen in yellow in the top half of the figure. Design courtesy of M. Fowler.

Figures 14, 21, 22 and 23 show the proposed layout in Hall C.

## I.3 Particle identification

In Fig. 24 the energy loss in the first layer of the CLAS scintillators is shown versus the velocity of the particle determined from the TOF. The energy loss is corrected for the light attenuation for each individual PMT. The position determination along the scintillator bars is determined using the time difference measured by the two PMTs on each scintillator. This is a standard CLAS procedure. The line shows the cut used to identify proton with $\beta < 0.9$.

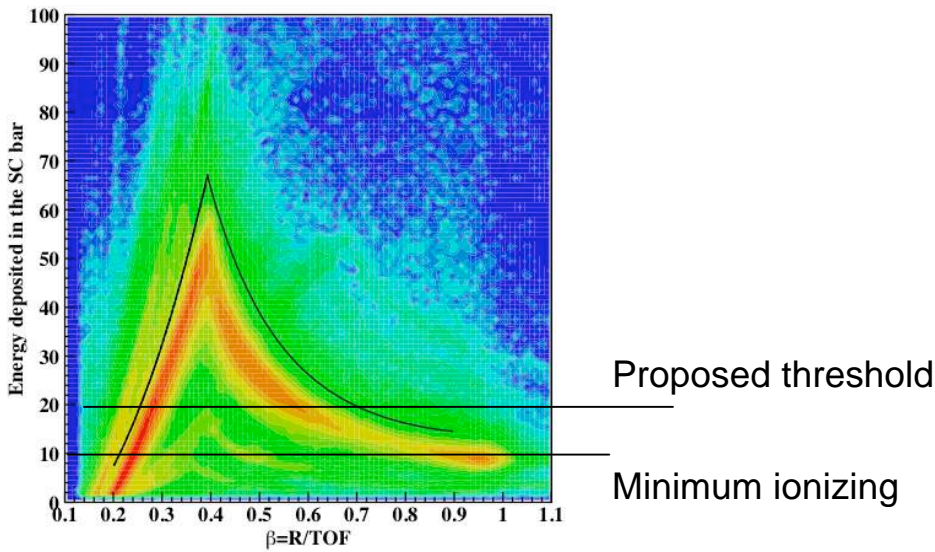

**Figure 24:** The energy loss in the CLAS scintillators versus the velocity $\beta$ as determined from the TOF.

Figure 25 shows the energy loss per cm in the scintillant and the TOF per meter of flight path as simulated using Geant.

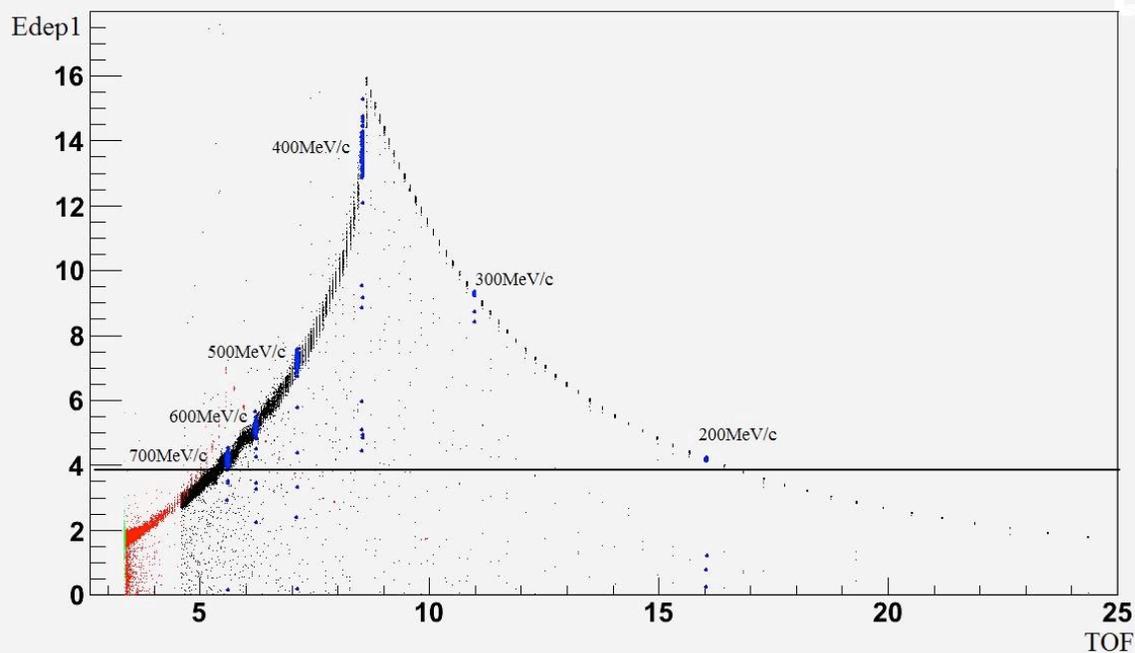

**Figure 25:** Energy loss versus TOF.

To reduce the large singles rates we will set the detector threshold in the first layer of scintillators for protons at 20 MeVee (twice the minimum ionizing energy deposit). We will set the detector threshold in the subsequent layers of scintillators for neutrons at 5 MeVee.

### I.4 Nucleon momentum resolution

The best momentum resolution is obtained by calculating the TOF per meter. From the CLAS operational experience, we know that the time resolution for the large angle counters is 200-250 ps. For detectors at about 5 meter from the target and 300-500 MeV/c nucleon

$$\frac{\Delta p}{p} = \frac{\Delta TOF}{TOF} = \frac{0.250 \text{ns}}{(50-33)\text{ns}} = 0.5 - 0.8\%$$

For recoil protons we can also determine their momentum using the energy loss. Based on experience with BigBite we expect a resolution of about 20 MeV/c.

$$\frac{\Delta p}{p} = \frac{20 \text{MeV}/c}{(500-300)\text{MeV}/c} = 4 - 7\%$$

## I.5 *x'*, *W'* resolution

The *x'* and *W'* are calculated based on the measurement of both the spectrometers and the LAD. The main goal of the experiment is to measure the ratio of cross sections at two different values of *x'* to extract the ratio of the struck bound nucleon structure functions.

The uncertainty in the determination of *x'* are presented below. The calculations are done with and without a GEM detector to improve the angular resolution of the scattered proton. For neutrons only the values without GEM are relevant.

| High *x'* kinematics | Low *x'* kinematics |
|---|---|
| $E' = 4.39 \pm 0.0022$ | $E' = 4.39 \pm 0.0022$ |
| $\theta_e = 17 \pm 0.057$ | $\theta_e = 13.5 \pm 0.057$ |
| $\|p_s\| = 0.35 \pm 0.0026$ | $\|p_s\| = 0.35 \pm 0.0026$ |
| $\theta_{sq} = 130 \pm 1.5$ (using GEM: 0.1) | $\theta_{sq} = 130 \pm 1.5$ (using GEM: 0.1) |
| $x' = 0.503 \pm 0.007$ (using GEM: 0.004) | $x' = 0.316 \pm 0.0046$ (using GEM: 0.003) |
| $W' = 2.18 \pm 0.024$ (using GEM: 0.01) | $W' = 2.518 \pm 0.02$ (using GEM: 0.009) |

## I.6 Neutron detection efficiency

For the neutron rates calculations, we assumed a detection efficiency of 5% per layer for the 300-500 MeV/c neutrons. The total detection efficiency in LAD is assumed to be 20%.

## I.7 Time Resolution

Neglecting the resolution in the measured time (better than 1 ns)

$$\frac{\Delta p}{p} = \frac{\Delta TOF}{TOF}$$

$$\Delta TOF = TOF \cdot \frac{\Delta p}{p} = (13.2 \text{ ns}) \frac{1}{\beta} \cdot \frac{\Delta p}{p} = (4\text{m})(3.3 \text{ ns/m})(E/p)\frac{\Delta p}{p} \approx (13.2 \text{ ns})(1 \text{ GeV})\frac{\Delta p}{p^2}$$

We assume that the uncertainty in the Energy-Loss determination of the momentum is 20 MeV/c. Then the largest time uncertainty is for the lowest momentum:

$$\Delta TOF = 13.2 \cdot \frac{0.020}{0.3^2} = 3 \text{ ns.}$$

We will assume a conservative resolving time of 5 ns.

Since the TOF is the only measurement of neutron momentum, the neutron resolving time is about 60 ns and depends on the full range from the highest to the lowest momentum accepted.

# Appendix II: Scattering chamber – conceptual design

The figure shows a conceptual design of a scattering chamber similar to the Hall A BigBite chamber, but adapted for Hall C. As in Hall C, the height of the chamber is 44.75 inches, the OD is 45 inches and the center of the beam hole is 24.25 inches above the bottom of the chamber. The thickness of the wall is 2 inches, which is the same as in Hall A. There will be flanges made for all openings which will also add strength to the chamber. Moveable vertical rods will provide additional support at selected locations. The actual window sizes and locations will depend on the final locations of the electron spectrometers and the LAD.

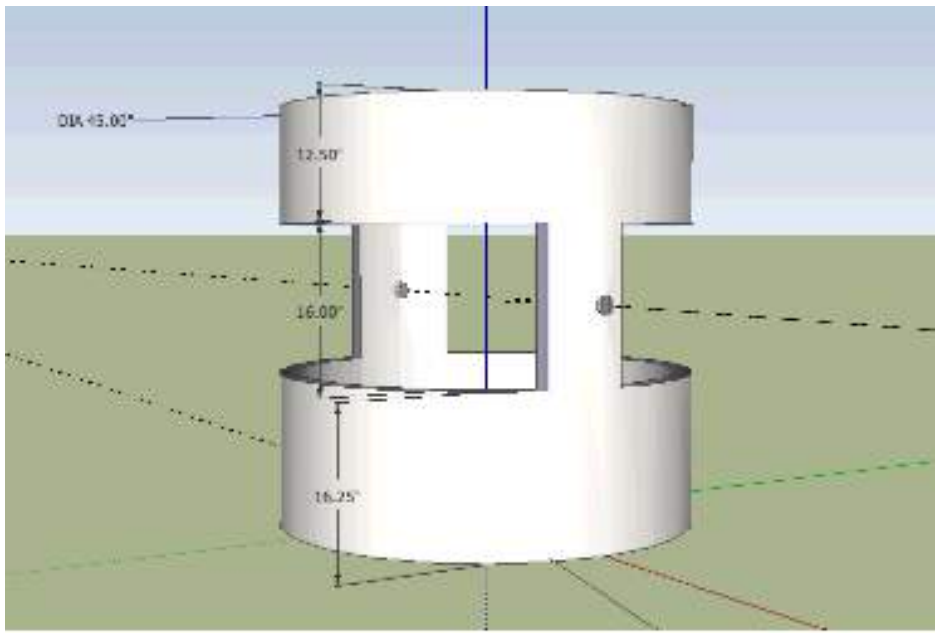